*Title*: Genomic evidence of rapid and stable adaptive oscillations over seasonal time scales in Drosophila

*Authors*: Alan O. Bergland[1*], Emily L. Behrman[2], Katherine R. O'Brien[2], Paul S. Schmidt[2], Dmitri A. Petrov[1]

*Affiliations*:
[1] Department of Biology, Stanford University, Stanford, CA 94305-5020
[2] Department of Biology, University of Pennsylvania, Philadelphia, PA, 19104
[*] Correspondence to: bergland@stanford.edu

*Running head*: Seasonal adaptation in Drosophila

*Abstract*: In many species, genomic data have revealed pervasive adaptive evolution indicated by the fixation of beneficial alleles. However, when selection pressures are highly variable along a species' range or through time adaptive alleles may persist at intermediate frequencies for long periods. So called 'balanced polymorphisms' have long been understood to be an important component of standing genetic variation yet direct evidence of the strength of balancing selection and the stability and prevalence of balanced polymorphisms has remained elusive. We hypothesized that environmental fluctuations between seasons in a North American orchard would impose temporally variable selection on *Drosophila melanogaster* and consequently maintain allelic variation at polymorphisms adaptively evolving in response to climatic variation. We identified hundreds of polymorphisms whose frequency oscillates among seasons and argue that these loci are subject to strong, temporally variable selection. We show that these polymorphisms respond to acute and persistent changes in climate and are associated in predictable ways with seasonally variable phenotypes. In addition, we show that adaptively oscillating polymorphisms are likely millions of years old, with some likely predating the divergence between *D. melanogaster* and *D. simulans*. Taken together, our results demonstrate that rapid temporal fluctuations in climate over generational time promotes adaptive genetic diversity at loci affecting polygenic phenotypes.

*Author summary*. Herein, we investigate the genomic basis of rapid adaptive evolution in response to seasonal fluctuations in the environment. We identify hundreds of polymorphisms (seasonal SNPs) that undergo dramatic shifts in allele frequency, on average moving in frequency between 40 and 60%, and oscillate between seasons repeatedly over multiple years likely inducing high levels of genome-wide genetic differentiation. We demonstrate that seasonal SNPs are functional, being both sensitive to an acute frost event and associated with two stress tolerance traits. Finally, we show that some seasonal SNPs are likely ancient balanced polymorphisms. Taken together, our results suggest that environmental heterogeneity can promote the long-term persistence of functional polymorphisms within populations that fuel fast directional adaptive response at any one time.

*Introduction.* All organisms live in environments that vary through time and such environmental heterogeneity can impose highly variable selection pressures on populations. In this situation, an allele may be beneficial during one environmental regime and subsequently deleterious during another. Such an allele would be subject to short bursts of directional selection, alternately being favored and disfavored. When this situation occurs in diploids, the heterozygote can have a higher geometric mean fitness than either homozygote and allelic variation at this locus could be maintained for long periods despite being subject to directional selection at any given time [1-8]. This situation is referred to as marginal overdominance and is a form of balancing selection.

Evidence for the maintenance of phenotypic and genetic variation by temporally variable selection has been observed in a variety of organisms. For instance, evolutionary response to rapid changes in selection pressures has been demonstrated for morphological and life-history traits in hamsters [9], finches [10], parrotlets [11], swallows [12], snails [13], ladybird beetles [14], freshwater copepods [15], midges [16], dandelions [17], foxes [18], aphids [19], spear winged flies [20,21], vinegar flies [22-24]) and others (reviewed in [25,26]. Chromosomal inversions and allozyme variants in a variety of drosophilids vary among seasons [27-33] suggesting that these polymorphisms confer differential fitness in alternating seasons. Further, in some species of drosophilids, life-history [34] [35]), morphological [36,37] and stress tolerance traits [38,39] vary among seasons suggesting that these traits respond to seasonal shifts in selection pressures.

Although theoretical models suggest that temporal variation in selection pressures can maintain fitness-related genetic variation in populations [1-8] and empirical evidence from a variety of species [9-39] demonstrates that variation in selection pressures over short time periods does alter phenotypes and allele frequencies, we still lack a basic understanding of many fundamental questions about the genetics and evolutionary history of alleles that undergo rapid adaptation in response to temporal variation in selection pressures. Specifically, do not know how many loci respond to temporally variable selection within a population, the strength of selection at each locus, nor the effects of such strong selection on neutral genetic differentiation through time. We do not know if adaptation at loci that respond to temporally variable selection is predictable. We do not know the relationship between loci that respond to temporally variable selection and spatially varying selection. Finally, we do not know whether temporal variation in natural selection promotes rapid adaptation at mutations that arise *de novo* or, rather, if it can lead to stable marginal overdominance at multiple loci and result in long-term maintenance of fitness-related genetic variation.

To address these questions, we estimated allele frequencies genome-wide from samples of *D. melanogaster* collected along a broad latitudinal cline in North America and in the spring and fall over three consecutive years in a single temperate orchard. We demonstrate that samples of flies collected in a single orchard over the course of several years are as differentiated as populations separated by 5-10° latitude. We identify hundreds of polymorphisms that are subject to strong, temporally varying selection and argue that genetic levels of genetic draft in the wake of rapid, multilocus adaptation is sufficient to explain the high degree of genetic turnover.We examine the genome-wide relationship between spatial and temporal variation in allele frequencies and find that spatial differentiation, but not clinality *per se*, in allele frequency is a good predictor of temporal variation in allele frequency and that northern populations are more 'spring-like' than southern ones. Next, we show that allele frequencies at SNPs subject to seasonal fluctuations in selection pressures become more 'spring-like' immediately following a hard frost event and that seasonally variably SNPs tend to be associated with two seasonally variable phenotypes, chill coma recovery and starvation tolerance. Finally, we demonstrate that some of the loci that respond to temporal variation in selection pressures are likely ancient, balanced polymorphisms predating the split of *D. melanogaster* from its sister species, *D. simulans*. Taken together, our results demonstrate, for the first time, that temporally variable selection can maintain fitness-related genetic variation at hundreds of loci throughout the genome for millions of generations if not millions of years.

*Results*

*Genomic differentiation through time and space*. To test for the genomic signatures of balancing selection caused by seasonal fluctuations in selection pressures, we performed whole genome, pooled resequencing of samples of male flies collected in the spring and fall over three consecutive years (2009-2011) in a temperate, Pennsylvanian orchard. We contrast changes in allele frequencies through time with estimates of allele frequencies we made from five additional populations spanning Florida to Maine along the east coast of North America (Fig. 1A, Supplemental Table 1). From each population and time point, we sampled approximately 50-100 flies and resequenced each sample to ~20-200X coverage (Supplemental Table 1). Estimates of allele frequency using this sampling design have been shown to be highly accurate [40].

As a point of departure and to provide context for understanding the magnitude of genetic variation through the seasons, we first examined genetic differentiation along the cline (Fig. 1B, Supplemental Fig. 1A). We calculated genome-wide average $F_{ST}$ between pairs of populations (excluding Pennsylvanian populations; hereafter 'spatial $F_{ST}$') as well as the probability that genome-wide average spatial $F_{ST}$ between pairs of populations is greater than expected by chance conditional on our sampling design and assuming panmixia using allele frequency estimates of 500,000 common polymorphisms (Supplemental table 1). Genome-wide average spatial $F_{ST}$ (Fig. 1B) as well as the probability that spatial $F_{ST}$ is greater than expected by chance (Supplemental Fig. 1A) is positively correlated with geographic distance ($r = 0.75$; $p = 7e-5$), a pattern consistent with isolation by distance [41]. Pooled resequencing did identify polymorphisms in or near genes previously shown to be clinal in North American populations (results not shown) demonstrating that clines are stable through time, suggesting that populations sampled along the cline represent resident populations, and further confirming that our pooled resequencing design gives accurate estimates of allele frequencies.

Next, we calculated genome-wide average $F_{ST}$ between samples collected through time in the Pennsylvanian population ('temporal $F_{ST}$') as well as the probability that genome-wide average temporal $F_{ST}$ is greater than expected by chance given our sampling design and assuming no allele frequency change through time (Fig. 1C, Supplemental Fig. 1B). Genome-wide average temporal $F_{ST}$ (Fig. 1C) as well as the genome-wide probability that observed $F_{ST}$ is greater than expected by chance (Supplemental Fig. 1B) increases with the difference in time between samples. The increase of temporal $F_{ST}$ with duration of time between samples increases non-linearly ($slope_{\text{log-log}} = 0.59$, $p_{\text{log-log slope=1}} = 0.0004$, $df=19$). Genome-wide average temporal $F_{ST}$ appears to asymptote by ~7 months, corresponding to the duration of time between fall samples and the subsequent spring sample. Remarkably, samples of the Pennsylvanian population collected one to three years apart are as differentiated as populations separated by 5-10° latitude, demonstrating high genetic turnover through time.

*Identification and genomic features of seasonal SNPs.* We sought to identify alleles whose frequency consistently and repeatedly oscillated between spring and fall over three years with the assumption that these polymorphisms would be the most likely to be adaptively responding to selection pressures that oscillate between the seasons. We identified seasonally variable polymorphisms using a generalized linear model (GLM) of allele frequency change as a function of season (spring or fall) that took into account read depth and the number of sampled chromosomes (see 'Materials and methods' for details).

Of the ~500,000 common SNPs tested, we identified approximately 1750 cyclically varying sites (hereafter 'seasonal SNPs') at FDR less than 0.3 (Fig. 2A, Supplemental Fig. 2A) that cycle approximately 20% in frequency between spring and fall. Changes in allele frequency of this magnitude correspond to selection coefficients of 5-50% per locus per generation (Fig. 2B, see Materials and Methods), assuming 10 generations per summer or 1-2 generations per winter. Given the statistical power of our experiment (Fig. 2B), we estimate there may be as many as 10 times as many sites that could cycle either directly in response to seasonally varying selection or could be linked to seasonal SNPs.

We note that our estimation of ~1750 seasonal SNPs should only be taken as a rough estimate of the number of seasonally varying SNPs: variance in linkage disequilibrium through the genome, heterscedasticity due to possible demographic events, statistical power, unbalanced sampling of flies and variance in read-depth among samples, modeling assumptions, and choice of FDR methodology will affect our ability to infer the exact number of seasonally varying SNPs. One way to address some of these issues (e.g., heteroscedasticity) is to model allele frequency change through time with generalized linear mixed-effect (GLMM) or general estimation equation (GEE) models that account, to varying degrees, for the structured, time-series nature of our data. Seasonal SNPs inferred with these models are highly congruent with seasonal SNPs inferred using a simple GLM (Supplemental Fig. 2D,E) and q-q plots of the distribution of p-values from GLM, GLMM and GEE models suggest that all GLM and GLMM modeling strategies fit the bulk of the genome well, with GEE models appearing to be anti-conservative (Supplemental Fig. 2B,C). However, the identification of a statistical excess of seasonally oscillating SNPs by any modeling strategy will be subject to a number of assumptions that will almost certainly be violated in some way or another and such violations could possibly lead to an increased false-positive rate. Because the false positive and false negative rates are unknown, we adopt an empirical strategy to demonstrate that the seasonal SNPs identified though a simple GLM are not a random sample of SNPs. Rather, these SNPs show many signatures consistent with natural selection relative to control SNPs that are matched for several biologically and experimentally relevant parameters such as chromosome, recombination rate, allele frequency, and SNP quality coupled with a rigorous blocked-bootstrap procedure that accounts for the spatial distribution of seasonal SNPs along the chromosome (see

Materials and Methods and Supplemental Table 3). We now proceed to demonstrate these enrichments.

Seasonal SNPs are enriched among functional genetic elements. These polymorphisms are likely to be in genic ($p = 0.054$) and coding regions ($p < 0.002$) and are enriched among synonymous ($p < 0.002$), non-synonymous ($p = 0.002$) and 3' UTR ($p = 0.024$, Fig. 2C) relative to control polymorphisms after controlling for the spatial distribution of seasonal SNPs along the chromosome using a block bootstrap procedure coupled with the identification of paired control SNPs matched for several key genomic features (Supplemental Table 3), such as recombination rate, average allele frequency in the Pennsylvanian orchard, chromosome, and SNP quality (see 'Block Bootstrap' section in Materials and Methods). Enrichment of adaptively oscillating polymorphisms among genetic elements suggests that these SNPs may affect organismal form and function through modification of protein function, translation rates, or mRNA expression and stability [42,43].

Next, we show that rapid shifts in allele frequency at seasonal SNPs perturb allele frequencies at nearby SNPs. Adaptively oscillating polymorphisms are in regions of elevated temporal $F_{ST}$ (Fig. 2D) and the elevation of temporal $F_{ST}$ decays, on average, by ~500bp, consistent with patterns of linkage disequilibrium in *D. melanogaster* [44]. Elevation of temporal $F_{ST}$ within 500bp of seasonal SNPs could contribute to high levels of genome-wide average $F_{ST}$ through time (Fig. 1C). Excluding SNPs within 500bp seasonal SNPs did not change patterns of genome-wide differentiation through time suggesting that genome-wide patterns of $F_{ST}$ through time are not driven by the seasonal SNPs themselves nor the SNPs in their immediate vicinity (Supplemental Fig. 3).

Seasonal SNPs are spread throughout the genome (Fig. 3A) and there is a 95% chance of finding at least one seasonal SNP per megabase of the euchromatic genome. In general, seasonal SNPs are not enriched among large, cosmopolitan inversions segregating in North American populations ($p > 0.05$, Supplemental Fig. 4), with only one inversion, *In3R(Mo)*, marginally enriched for seasonal SNPs ($p = 0.02$, with $p = 0.18$ after Bonferroni correction for multiple testing). In addition, seasonal SNPs are significantly more common in the Pennsylvanian orchard population than polymorphisms perfectly linked [45] to large cosmopolitan inversions (Fig. 2E) and polymorphisms linked to inversions do not vary between seasons (Fig. 2E, $p > 0.05$), including those linked to *In3R(Mo)*. Therefore, enrichment of seasonal SNPs within *In3R(Mo)*, if present, is most likely due to increased linkage disequilibrium caused by decreased recombination surrounding this inversion [46]. Taken together, these results demonstrate that the inversions themselves do not cycle seasonally in the Pennsylvanian population in any appreciable manner (Fig. 2E) and suggest that adaptive evolution to seasonal variation in selection pressures is highly polygenic.

*Relationship between spatial and temporal variation in allele frequencies*. To test the hypothesis that spatially varying selection pressures along the latitudinal cline reflect seasonally varying selection pressures in the Pennsylvanian population, we examined the relationship between temporal and spatial variation in allele frequencies. To quantify spatial variation in allele frequency, we calculated two statistics. First, we estimated average pairwise $F_{ST}$ among all populations for each SNP ('spatial $F_{ST}$'). Second, we estimated clinality for each SNP by calculating the q-value ('clinal *q*-value') of the relationship between allele frequency and latitude using a generalized linear model that takes into account read depth and the number of sampled chromosomes. Spatial $F_{ST}$ and clinal *q*-value are highly correlated ($r = 0.63$, $p < $ 1e-10; Supplemental Fig. 5) demonstrating that most spatial variation along the latitudinal cline is represented by monotonic changes in allele frequency between northern and southern populations.

We calculated the number of clinally varying polymorphisms (clinal q-value < 0.1) and the number of adaptively oscillating polymorphisms per common segregating SNP (average, North American MAF > 0.15) per megabase of the genome (Fig. 3A). Approximately one out of every three common polymorphisms varies with latitude with FDR < 0.1 whereas only one out of every three thousand polymorphisms varies predictably between seasons with FDR < 0.3 (Fig. 3A). Although our ability to detect clinal SNPs at FDR < 0.1 is greater than our ability to detect seasonal SNPs at FDR < 0.3 (Supplemental Fig. 6), differences in power cannot explain the three order of magnitude difference in the expected number of clinal and seasonal SNPs (cf. Fig. 2B, Supplemental Fig. 6). Therefore, different evolutionary forces such as demography and selection pressures that are clinal but not seasonal, shape allele frequencies through time and space.

Next, we formally tested whether seasonal SNPs are enriched among spatially varying SNPs. Spatially varying SNPs, as defined by spatial $F_{ST}$, are more likely to be seasonal SNPs than expected by chance (Fig. 3B), and the odds of this enrichment increases with increasing spatial differentiation. However, there is no enrichment of seasonal SNPs among clinal SNPs as defined by clinal *q*-value (Fig. 3C).

Because of the relationship between spatial differentiation and seasonal variation in allele frequencies (Fig. 3B) and because of parallels between spatial and seasonal variation in climate, we hypothesized that northern populations should be more spring-like and southern populations should be more fall like in allele frequencies at the seasonal SNPs. To test this hypothesis, we calculated the absolute difference in allele frequencies for each population sampled along the cline with the average spring and fall allele frequency estimates for the Pennsylvanian population for all seasonal SNPs. Indeed, high latitude populations are more similar to spring Pennsylvanian populations and those from low latitude are more similar to fall populations (Fig. 3D) demonstrating that latitudinally varying selection pressures at least partially reflect seasonally varying selection pressures.

*Immediate adaptive response to an acute frost event*. In the late fall of 2011, about two weeks after our 2011 'fall' sample was collected, a hard frost occurred in the Pennsylvanian orchard (Fig. 4A). We were able to obtain a sample of *D. melanogaster* approximately one week after the frost and we estimated allele frequencies genome-wide from this sample. We hypothesized that allele frequencies at seasonal SNPs would predictably change following the frost event and would become more 'spring-like.' To test this hypothesis, we calculated the probability that the post-frost allele frequency at seasonal SNPs overshoot the long-term average allele frequency (i.e., become more spring-like). We also estimated this probability for control polymorphisms, matched to adaptively oscillating polymorphisms by several characteristics (Supplemental table 3) including, importantly, difference in allele frequency between the long-term average and the pre-frost allele frequency. This later control is essential given that some shift in the spring-like direction is expected here simply by chance from the regression to the mean.

The probability that seasonal SNPs overshoot the long-term average allele frequency is ~43%, whereas ~35% of control polymorphisms overshoot the long-term average. The significant ($\log_2$(OR) = 0.48, $p < 0.002$) excess of adaptively oscillating polymorphisms that become more spring-like following the frost event demonstrates that these SNPs respond to acute changes in climate and that cold temperatures associated with winter is a primary selective force acting on this population shaping allele frequencies between seasons.

*Association with seasonally variable phenotypes.* Chill-coma recovery time and starvation tolerance are two phenotypes that vary seasonally in drosophilid populations [47-52]. Accordingly, we hypothesized that the winter-favored allele at seasonal SNPs would be associated with decreased chill-coma recovery time and increased starvation tolerance. To test this hypothesis, we used allele frequency data from tail based mapping of chill-coma recovery time and starvation tolerance [53]. We show that the winter favored allele at seasonal SNPs is more likely to be associated with fast chill coma recovery time than expected by chance across a range of GWAS p-values (Fig. 5A). A similar analysis of starvation tolerance was equivocal but the general pattern is that the winter-adaptive allele is associated with increased starvation tolerance (Fig. 5B).

*Long term balancing selection.* Balancing selection caused by variation in selection pressures through time can in principle maintain allelic variation at adaptively oscillating loci and elevate levels of neutral diversity surrounding these balanced polymorphisms. Thus, if seasonal variation in selection pressures promotes balanced polymorphisms we hypothesized that seasonal SNPs would be old and in regions of elevated polymorphism.

We tested the hypothesis that seasonal SNPs are old by first examining their allele frequencies in a broad survey of African *D. melanogaster* populations [54]. Approximately 5% of seasonal SNPs are rare in Africa (MAF < 0.01), however these SNPs are not more likely to be rare in Africa than control polymorphisms ($\log_2$(odds ratio) = 0.96; $p = 0.328$). Because the vast majority of seasonal SNPs segregate in Africa, it appears that adaptation to temperate environments occurred mostly from old, standing genetic variation.

Balancing selection acts to maintain alleles at intermediate frequencies for long periods of time and, in some instances, can maintain polymorphism across species boundaries [55,56]. We examined whether seasonal SNPs showed signatures of long-term balancing selection by examining patterns of polymorphism surrounding orthologous regions in *D. simulans*, the sister species to *D. melanogaster*. We note that the following analyses are conservative because we underestimate *D. simulans* diversity given the small number (< 6) of *D. simulans* haplotypes used.

First, we demonstrate that seasonal SNPs are approximately 1.5 times more likely to be polymorphic and share the same two alleles in both species relative to control SNPs. This pattern is observed for all seasonal SNPs (Fig. 6A, $p < 0.002$) and for seasonal SNPs residing in genes (Fig. 6A, $p < 0.002$). The increased probability of shared polymorphism between *D. melanogaster* and *D. simulans* at seasonal SNPs could, in principle, be driven by an over-representation of synonymous, genic SNPs (Fig. 2C). Unless synonymous SNPs are in four-fold degenerate positions, certain mutations may cause them to be non-synonymous thereby limiting the number of possible neutral allelic states and increasing the probability of shared polymorphism between species. However, adaptively oscillating SNPs that do not reside in synonymous sites are also more likely than expected by chance to be polymorphic and share the same two alleles by state in *D. melanogaster* and *D. simulans* (Fig. 6A, $p = 0.014$).

The co-occurrence of shared polymorphism between *D. melanogaster* and *D. simulans* could result from three evolutionary mechanism. First, trans-specific polymorphisms could result from adaptive introgression. This scenario seems implausible given the high degree of pre- and post-zygotic isolating mechanisms between these two species [57,58]. Furthermore, if trans-specific polymorphisms resulted from recent adaptive introgression we would expect average pairwise divergence between *D. melanogaster* and *D. simulans* surrounding seasonal SNPs to be smaller than at control SNPs. However, there is no significant difference in estimates of divergence between seasonal and control SNPs ($p = 0.7$ for windows ±250bp). Second, trans-specific polymorphisms could result from convergent adaptive evolution. Finally, trans-specific polymorphisms could be millions of years old [59], predating the divergence of *D. melanogaster* from *D. simulans*. While we cannot differentiate these latter two mechanisms, the most parsimonious explanation is that trans-specific seasonal SNPs predate the divergence of these two sister species.

*Discussion*

Herein, we present results from population based resequencing of samples of flies collected along a latitudinal cline in North America and over three years during the spring and fall in a Pennsylvanian orchard. We identify repeatable and dramatic changes in allele frequencies through time at hundreds of polymorphisms spread throughout the genome. Response to strong selection at these seasonal SNPs likely drives genetic differentiation through time at linked, neutral polymorphisms. This process leads to genome-wide differentiation between samples collected several years apart comparable to populations separated by 5-10° latitude. Seasonal SNPs are likely to be functional as they show enrichment at functional sites, vary predictably among populations sampled along the cline, respond immediately to a hard frost event, and are associated with phenotypes previously shown to vary seasonally in temperate *D. melanogaster* populations. Finally, our results suggest that some adaptively oscillating SNPs are possibly millions of years old, predating the split of *D. melanogaster* from its sister species *D. simulans*. Taken together, our results provide the first genomic picture of balancing selection caused by temporal fluctuations in selection pressures and provide novel insight into the biology of marginal overdominance.

*The plausibility of seasonally variable selection*. We have argued that adaptive response to seasonally fluctuating selection at no less than 25-50 loci is necessary to generate the high levels of genome-wide genetic differentiation through time observed in the Pennsylvanian population. Next, we considered the plausibility of such strong selection and estimated the upper bound of the number of loci that could independently respond to seasonally variable selection. To do so, we modeled independent selection at 1-10,000 simulated seasonal SNPs whose allele frequency change was drawn from the observed allele frequency change at seasonal SNPs. Using a simple Poisson model (see Materials and Methods), we estimated the minimum fall census size required for that number of loci to shift in allele frequency during one or two rounds of truncation selection. Using these models, we sought to estimate the most likely number of seasonal SNPs that could independently respond to seasonally variable selection by contrasting model-based estimates of population size with our best estimates of population size in the field.

Although fall census size of *D. melanogaster* in the focal Pennsylvanian population is unknown, some estimates of drosophilid population size have been made. Global population size of *D. melanogaster* is likely to be extremely large, greater than $10^8$ [61]. However, estimates of local population size made from mark-release-recapture methods report census sizes on the order of $10^4$ to $10^5$ [62-64]), with considerable variation among seasons, years and locales. *D. melanogaster* samples from orchards and vineyards often exceed $10^4$ flies [65,66] and thousands of flies can easily be collected over large compost piles (Bergland pers. obs.). Therefore, we speculate that census size of temperate *D. melanogaster* populations at any locale is a function of the local ecology (e.g., amount of windfall fruit, number and size of compost piles, humidity) and given the favorable conditions in the focal Pennsylvanian orchard (Schmidt pers. obs.), large census sizes of more than $10^5$ are conceivable. If fall census size in the Pennsylvanian population is on the order of $10^5$, our truncation selection model suggests that no more than several hundred (200-700, Fig. 7C) seasonal SNPs could respond to seasonally varying selection independently. We note that increasing the number of generations of winter-like selection pressures or the fall census size would lead to a concomitant increase in the number of seasonally selected loci that could independently respond to seasonally varying selection pressures.

Our survey of temporal changes in allele frequency identified 1750 seasonal SNPs. Unless local census size in the Pennsylvanian population were unrealistically large – on the order of $10^{10}$ or $10^{20}$ – it is likely that not all of these loci could respond to selection independently. Our model suggests, however, that a large fraction, on the order of 200-700 could vary independently in every cycle. One explanation for cycling in the remaining SNPs is linkage with loci responding to seasonally variable selection. It is possible that this linkage is generated either stochastically and neutrally or, alternatively, by selective processes such as assortative mating [67] or epistatic selection [68,69]. For instance, if winter adapted flies were more likely to mate with other winter adapted flies during the summer, winter adapted alleles may become coupled and linkage disequilibrium between these alleles could increase. Similarly, certain forms of epistatic interactions could also generate linkage disequilibrium between seasonal SNPs if, for instance, couplings of winter and summer favored alleles at multiple loci were particularly deleterious relative to winter-winter or summer-summer combinations. The net effect of selective mechanisms that promote positive linkage disequilibrium between seasonal SNPs is that the effective number of 'independently' seasonally selected loci decreases. If seasonal SNPs are in linkage disequilibrium due to selective processes, it would imply that more than 200-700 seasonal SNPs contribute to organismal form and function and modify fitness during the summer and winter.

*Seasonally variable selection is sufficient to generate patterns of allele frequency change through time*. Despite empirical support for the conclusion that seasonal SNPs show many signatures consistent with adaptive response to seasonally variable selection, drift, caused by cyclic population booms and busts, or migration from neighboring demes are alternative mechanisms that could drastically perturb allele frequencies in the Pennsylvanian population and could generate some of the genome-wide patterns we observe. We address these possibilities here and conclude that neither cyclic changes in population size nor seasonal migration can plausibly explain the extent of genome-wide genetic differentiation through time, the observed number of seasonal SNPs, nor the enrichment of seasonal SNPs among many distinct genomic features (e.g., Figs. 2-6). In contrast, we show here through several simulation approaches that rapid adaptive evolution to seasonal fluctuations in selection pressure is sufficient to explain patterns of allele frequency change through time. Furthermore, we discuss how large scale migration is internally inconsistent with certain aspects of our data. Taken together, we conclude that rapid adaptive evolution to seasonally variable selection is sufficient to explain the patterns of allele frequency change through time at seasonal SNPs and at linked neutral loci that we observe in our dataset.

First, we assessed the possibility that extensive drift caused by population contraction every winter [31,70,71] could generate genome-wide patterns of genetic differentiation through time observed in our data. To do so, we conducted forward genetic simulations that model biologically plausible variation in population size and included loci that cycle in frequency due to variable selection pressures [72]. For these simulations, we modeled a 20Mb chromosome with constant recombination rate of 2cM/Mb, representing the genome-wide average recombination rate in *D. melanogaster* [73]. We simulated population contraction to one of various minimum, 'overwintering' population sizes followed by exponential growth over 10 generations in the 'summer' to a fixed maximum population size. In these models, we included various numbers of loci that respond to seasonally varying selection. Selection coefficients for each locus were set such that allele frequencies at selected sites oscillated by ~20% between 60 and 40%, representing the average change in allele frequency we actually see between spring and fall at seasonal SNPs. Finally, we placed 500 neutral loci randomly along the simulated chromosome and measured $F_{ST}$ at these neutral loci between three 'spring' (i.e., first generation of population expansion) and 'fall' (last generation of population expansion) samples. See Materials and Methods for more details these models.

In the absence of seasonal selection, these forward simulations suggest that overwintering $N_e$ would have to be exceedingly low (~20) to generate levels of $F_{ST}$ between spring and fall as high as we observe in our data (Fig. 1C). However, with overwintering $N_e$ of 200 and 5-10 seasonally adaptive SNPs per chromosome arm, simulated $F_{ST}$ at neutral loci is on the order of 0.002, which we observe in our data (Supplemental Fig. 1B). While we do not know overwintering population size, we speculate it could be on the order of 200 flies or more [70,71] and conclude that at least 25-50 (5-10 per main chromosome arm) loci are sufficient to generate patterns of differentiation we observe through time. Note that increasing the overwintering population size requires concomitant increase in number of seasonally selected loci.

We regard overwintering population sizes of ~20 flies to be inconsistent with certain aspects of our data and also implausible given what we know about the biology of the species. First, such a severe population contraction would result in reduction of genetic diversity, particularly for low frequency alleles. However, the observed allele frequency spectrum between fall and the following spring samples is similar and spring samples do not exhibit the expected loss of low frequency polymorphisms that would result from a population contraction to 20 individuals (Supplemental Fig. 7). Second, population contraction to 20 individuals would often lead to population extirpation in the Pennsylvanian orchard and would certainly lead to extirpation at localities further north that experience more severe winters. However, *D. melanogaster* are routinely collected in Northern orchards very early in the season [74] and are routinely found in populations at as far north as 45° (Schmidt pers. obs). Furthermore, certain rare alleles have persisted in northern *D. melanogaster* populations for upwards of 30 years [74] *cf* [76] and allele frequency clines are relatively stable over decadal scales [77] demonstrating that high latitude populations are not frequently extirpated and that overwintering bottlenecks cannot be so severe as our neutral simulations would require.

In our forward simulations, seasonally variable selection is sufficient to generate high levels of genome-wide genetic differentiation through time. In addition, our forward simulations are consistent with the increase of genome-wide average $F_{ST}$ through time excluding polymorphisms that are within 500bp of seasonal SNPs (Supplemental Fig. 3). In our simulations, 500 neutral loci were placed randomly along a 20Mb chromosome and were initially completely unlinked to selected loci. Therefore, the high levels of simulated $F_{ST}$ are a consequence of genetic draft acting over long physical distances with low to moderate linkage disequilibrium between neutral and selected polymorphisms. Our observation that genome-wide average $F_{ST}$ (excluding polymorphisms near seasonal SNPs, Supplemental Fig. 3) increases with time resembles our simulations suggesting that draft can perturb allele frequencies over long genetic distances.

We also note that long-range genetic draft, caused by rapid adaptation of old alleles to seasonally variable selection would likely cause an asymptotic change in genome-wide temporal $F_{ST}$, whereas a purely drift based model would likely cause a monotonic increase in genome-wide $F_{ST}$ through time. Seasonal SNPs tend to be old and are therefore likely found on a diverse array of haplotypes. Therefore, the exact composition of haplotypes that rise and fall every seasonal cycle will be somewhat stochastic giving rise to a high genome-wide $F_{ST}$ over a duration of time less than ~7 months (the duration of time between fall and the following spring). Among years, genome-wide average $F_{ST}$ would possibly plateau if local $N_e$ were large (as we suspect it is, see *Discussion: The plausibility…*), coupled with the effects of recombination, gene conversion, and low-level migration from neighboring demes or populations. In contrast, a purely drift based model of an isolated deme would display monotonic increase in genome-wide $F_{ST}$ through time.

Next, we explore the possibility that migration could drastically alter allele frequencies in the Pennsylvanian population and generate the large number of loci that vary repeatedly among seasons. First, we examined a simple but general demographic model where the Pennsylvanian orchard population becomes extirpated every year and recolonized from a refugium such as a southern population or a large, local site such as a compost pile. Either situation is plausible given the purportedly high rates of migration in North American *D. melanogaster* populations [78] and what little is known about the overwintering biology of high latitude *D. melanogaster* [75]. In our model, we envisioned a resident, refugal population with stable allele frequencies across years that colonizes the orchard population. In this model, the orchard would be colonized in the early season with a random selection of flies from the refugium and would therefore have aberrant allele frequencies. As more migrants arrived to the orchard from the refugium, allele frequencies at the orchard would stabilize to that of the source population. In such a scenario, allele frequencies in spring samples could vary considerably but a small fraction of SNPs would, by chance, have the same aberrant allele frequencies year after year.

We calculated the expected number SNPs that would vary in a repeatable way by chance alone as a function of the number of initial migrants (Fig. 7B). For instance, if five migrants arrived at the orchard prior to our spring sample every year, approximately 1300 SNPs would vary repeatedly among seasons producing similar patterns to the observed change in allele frequency through time as at 'seasonal SNPs' (Fig. 2A). However, if four migrants arrived at the orchard prior to our sampling, ~2600 SNPs would vary repeatedly but if six migrants arrived, only ~700 would. Although the expected number of sites that oscillate under this migration model with 5 migrants is approximately the number we observe, we note that the expected number is highly dependent on the exact number of migrants. It seems unlikely that exactly five flies would migrate from the refugium to the orchard before our first spring sample three times in a row. Therefore, the extreme sensitivity of the expected number of sites to the number of migrants makes this general demographic scenario implausible. We are therefore led to conclude that the simple migration model presented here is likely insufficient to explain changes in allele frequency through time in the Pennsylvanian orchard.

In addition to our conclusion that a simple model of recolonization of the orchard is insufficient to explain the number of seasonally variable loci we observe, our data are also inconsistent with large-scale migrations from adjacent populations. For instance, if a large-scale migration from the South to resident northern populations were to occur, we would expect that clinally varying SNPs should also vary seasonally. Such a pattern would be expected both if large scale migration occurred randomly or were genotype dependent. However, seasonal SNPs are not enriched among clinally varying polymorphisms (Fig. 3C). Therefore, we conclude that large-scale migration does not play a major role shaping seasonal variation in allele frequencies in the Pennsylvanian orchard. Furthermore, even if seasonal SNPs were enriched among clinally varying polymorphisms (which they are not), adaptation to seasonally variable selection would need to be invoked in order to explain the yearly shift in allele frequencies every winter.

Taken together, models presented here demonstrate that seasonal boom-bust or migration-based scenarios are insufficient to explain allele frequency change through time in the Pennsylvanian population. While temperate populations of *D. melanogaster* clearly undergo cyclic population booms and busts due to changes in climate associated with the season, the extent of these population contractions necessary to generate the patterns of genetic variation through time that we observe would be too extreme to allow for stable population persistence. Similarly, the Pennsylvanian population certainly exists as a part of a complex metapopulation and experiences immigration and emigration. However, analysis of a simple demographic model of population recolonization during the spring is also insufficient to explain the patterns of allele frequency change through time that we observe and our data are internally inconsistent with a model of large-scale migration from the South. Finally, we point out that the boom-bust and recolonization models we presented here are clearly oversimplifications and that there are other, more complex demographic models that we have not explored. Nonetheless, any stochastic demographic event would affect SNPs throughout the genome with equal probability. Many aspects of our data clearly show that seasonal SNPs are not random but rather show signatures consistent with both functional effect and long-term balancing selection such as enrichment in specific classes of genetic elements, association with seasonally variable phenotypes and predictable shifts in allele frequency in response to acute and persistent variation in climate. Therefore, while we cannot conclusively rule out the possibility that demographic events affect the temporal dynamics of allele frequencies at seasonal- and non-seasonal SNPs in the Pennsylvanian population, these demographic events are most likely coupled with adaptive evolution in response to temporally varying selection pressures.

*Functional properties of adaptively oscillating polymorphisms*. Temperate populations of *D. melanogaster* are exposed to high levels of environmental heterogeneity among seasons due to changes in various aspects of the environment including temperature, humidity, and nutritional quality and quantity. These shifts in the environment are primary determinants of cyclic population booms and busts [62,63,75] and impose strong temporally and spatially variable selection. Intuition, theoretical models [79], laboratory experimentation [35], and inference from patterns of clinal variation [80-82] and seasonal variation in morphological, behavioral and life-history traits suggest that alternate seasons favor differing life-history strategies. In general, populations exposed to more harsh conditions such as those from Northern locales or those collected early in the season are larger [83,84], more stress tolerant [48-50,82], longer lived [81], and are less fecund [81,85] than those collected in Southern locales or during the fall. The general picture that emerges, therefore, is that in temperate populations winter conditions select for hardier but less fecund individuals whereas summer selects for high reproductive output at the cost of somatic maintenance. Nonetheless, there is surprisingly little evidence directly linking adaptive differentiation between seasonally favored genetic polymorphisms, phenotypes and environmental perturbations (but see [35]). Herein we present several key results that directly link seasonal and spatial patterns of genotypic and phenotypic variation and with environmental perturbations.

    First, we demonstrate that acute bouts of cold temperature elicit adaptive response at seasonally oscillating polymorphisms (Fig. 4). Heretofore, the specific environmental factors altering allele frequencies through time and space among dipteran species has generally remained elusive largely stemming from the fact that many aspects of the environment co-vary over temporal and spatial scales. Here we show that acute exposure to sub-freezing temperatures in the field shifts allele frequencies in a spring like direction at seasonal SNPs but not at control polymorphisms, thereby demonstrating that sharp modulation of temperature can act as a selective force in the field. While post-frost allele frequencies at seasonal SNPs move in a spring-like direction, they do not reach average spring allele frequencies. This suggests that multiple frost events, long-term exposure to cold temperatures or other factors linked to winter conditions such as starvation also impose strong selection in temperate populations.

    Next, we demonstrate that environmental differences among populations predict changes in allele frequency at seasonal SNPs. Environmental factors that vary over seasonal time scales also vary with latitude. This fact has facilitated studies that substitute space for time and has led to a paradigm in many aspects of contemporary research in drosophilid evolutionary ecology of examining phenotypic and genetic differentiation along latitudinal (and altitudinal) clines as a proxy for studying adaptation to temperate

environments [86]. Using allele frequency estimates that we made from populations sampled along the North American latitudinal cline, we demonstrate that southern populations are more 'fall-like' at seasonal SNPs whereas northern populations are more 'spring-like' (Fig. 3D). Northern populations experience more severe winters and have shorter growing seasons; therefore, we speculate that the changes in allele frequency at adaptively oscillating polymorphisms along the cline is because (1) the summer favored allele would be at lower frequency due to stronger selection during the winter and (2) the summer favored allele would not rise in frequency as much during the summer because of the shorter growing season. The converse would be the case for Southern populations.

Finally, we provide a direct connection between seasonal SNPs and ecologically relevant phenotypic variation. Previous studies have demonstrated that two important stress tolerance traits, chill coma recovery time and starvation resistance vary in predictable ways among temperate populations of *D. melanogaster*. Northern populations tend to have fast chill coma recovery time [87-89] recapitulating deeper phylogenetic patterns among drosophilids originating from temperate and tropical locales [47]. Evidence for latitudinal variation in starvation tolerance is more equivocal with low latitude populations of *D. melanogaster* being more starvation tolerant in some studies but not significantly so in others [48,90] and closely related species showing equally ambiguous patterns [52,90,91]. However, diapause competent genotypes that are at high frequency in Northern populations and in the spring show increased starvation tolerance [51] suggesting that spatial and temporal differentiation in starvation tolerance may be parallel in the context of specific polymorphisms. Nonetheless, because selection pressures along latitudinal clines are generally parallel with seasonal selection pressures (e.g., Fig. 3D) we reasoned that winter adapted alleles at seasonal SNPs would be associated with fast chill coma recovery time and increased starvation tolerance.

We show that winter adapted alleles at seasonal SNPs are likely to be associated with fast chill coma recovery time and, to a lesser extent, starvation tolerance (Fig. 5). The strength of the relationship between seasonal SNPs with these two phenotypes likely differs for many reasons, including intrinsic differences in the statistical power and the complex genetic architecture of these traits. Nonetheless, that seasonal SNPs are associated with chill coma recovery and starvation tolerance in the predicted direction given our prior knowledge of seasonal variation in these two traits strongly suggests that seasonal SNPs are functional and affect seasonally dependent fitness via stress tolerance traits. In addition, the concordance between seasonal SNPs and SNPs moderately associated with chill coma recovery time and starvation tolerance suggests that the intermediate frequency SNPs that we are investigating here have small effects on phenotype compared to low-frequency, possibly deleterious alleles but nonetheless have large effects on average population fitness.

Taken together, our analysis has directly linked adaptive oscillations at hundreds of polymorphisms in *D. melanogaster* to specific and persistent differences in climate and to phenotypes known to be under diversifying selection through time and space. Our results support the hypothesis that stress tolerance traits are favored during the winter and disfavored during the summer. Stress tolerance traits such as chill coma recovery time and starvation tolerance often have negative genetic correlations with reproductive output [51,92] or development time [93], two phenotypes that would be favored during exponential growth during the summer. Therefore, it is likely that a subset of seasonal SNPs directly contribute to a tradeoff between stress tolerance and reproductive output.

Because *D. melanogaster* originated in sub-Saharan Africa and colonized the world in the wake of human migration 200-10,000 years ago [94] it has been hypothesized [95] that phenotypes favored during the winter are derived whereas those favored during the summer are ancestral with respect to tropical, African populations. Although we show that the vast majority of seasonal SNPs are common in Africa, a small set (~ 5%) are rare, segregating at less than 1%. Somewhat surprisingly, summer favored alleles are more likely to be rare in Africa than winter favored alleles ($\log_2$(odds ratio) = 0.475; $p = 0.018$) suggesting that some environmental aspects of summer in temperate orchards are new for *D. melanogaster*. Consistent with the observation that flies sampled at low latitudes are likely subject to intense intra- and inter-specific competition [83], we speculate that the cornucopia of rotten fruit during the summer in mid- to high-latitude locales coupled with decreased inter-specific competition [96] is a novel environment that has allowed formerly rare alleles associated with increased reproductive output to flourish.

*Long-term, polygenic balancing selection and ecological generality*. Herein, we present several lines of evidence demonstrating that hundreds of loci adaptively respond to seasonal fluctuations in the environment. Despite (or because of) the fact that these loci promote rapid adaptive evolution, they have remained polymorphic for millions of generations within *D. melanogaster* and many possibly predate the divergence of *D. melanogaster* and *D. simulans* ~5 million years ago. Taken together, these observations suggest that alleles at these loci have been maintained by environmental heterogeneity for exceptionally long periods of time. Long-term balancing selection is typically regarded as an evolutionary oddity, found predominantly in the genetic systems regulating host-pathogen interactions, self-incompatibility, and sex-determination [55,97]. Here, we show for the first time that environmental heterogeneity promotes long-term balanced polymorphisms at hundreds of loci that affect quantitative, stress tolerance traits.

The long-term persistence of these adaptively oscillating polymorphisms across populations, continents, and species suggests that these polymorphisms contribute to short-term and local adaptation in response to very generalized environmental conditions. This is in contrast to the hypothesis [95] that adaptation to temperate environments in *D. melanogaster* was largely in response to novel environments, exclusively associated with life in northern, temperate locales. Rather, we speculate that the selective pressures associated with seasons in temperate environments are merely manifestations of general selective pressures that are natural consequences of cyclic population booms and busts. That is, during times of plenty, such as during the summer in temperate locales, populations rapidly expand and alleles that confer increased reproductive output or faster time to sexual maturity are strongly favored. However, when population size contracts due to biotic and abiotic stressors such as those experienced during winter, alleles that confer increased stress resistance are favored.

Cyclic population booms and busts are almost certainly a perennial feature of *D. melanogaster* populations, are a likely common occurrence in highly fecund species that exploit ephemeral resources, and may be an inherent property of species in general [98]. If true, we speculate that such species may harbor alleles that promote reproductive fitness during population growth (at the cost of somatic maintenance) and increase stress tolerance (at the cost of reproductive growth) during population contraction. Such balanced polymorphisms may be particularly common for species whose population cycles are decoupled from predictable environmental cues (e.g., photoperiod) but are rather linked to stochastic changes in resource abundance. For species such as these, including many microorganisms and invertebrates, balanced polymorphisms maintained by environmental heterogeneity through time and space may be the norm rather than the exception.

*Materials and methods*

*Fly collections*. We resequenced samples of *D. melanogaster* from populations spread along a broad latitudinal cline in North America and during multiple time points over three consecutive years (2009 to 2011) at the Linvilla Orchard in Media, PA (39.9°N, 75.4°W). From each locality and sampling period, we collected ~50-200 *D. melanogaster* largely by aspiration from individual fruits or baiting at strawberry fields and apple and peach orchards, established isofemale lines and collected male progeny at generation 1-5 for sequencing. One male progeny per isofemale line per population was pooled together to generate template DNA for high throughput sequencing (Supplemental table 1). The only two exceptions are the second replicate sample from Maine which was derived from wild-caught males and the sample from North Carolina which was sampled from the DGRP inbred lines. For the DGRP population, we resequenced a pooled sample consisting of one male from each of 92 DGRP strains and used allele frequency estimates from pooled samples when estimating clinality (see [40] for more information on this sample and [44] for more information on this population).

*Sequencing and bioinformatics of pooled samples*. DNA libraries were prepared for sequencing on the Illumina HiSeq2000 platform following standard practices. Raw, paired-end 100bp sequence reads were mapped to the *D. melanogaster* reference genome version 5.39 using *bwa* version 0.5.9-r16 [99] allowing for a maximum insert size of 800bp and no more than 10 mismatches per 100bp. PCR duplicates (~5% per library) were removed using *samtools* version 0.1.18 [100] and local realignment around indels was performed using GATK version 1.4-25 [101]. We mapped SNPs and short indels (i.e., those occurring within the sequence reads) using CRISP [102], excluding reads with base or mapping quality below 10. SNPs mapping to repetitive regions such as microsatellites and transposable elements, identified in the standard RepeatMasker library for *D. melanogaster* (obtained from http://genome.ucsc.edu) were excluded from analysis as were SNPs within 5bp of polymorphic indels. SNPs with average minor allele frequency less than 15%, with minimum per-population coverage less than 10X or maximum per-population coverage greater than 400X were removed from analysis. Finally, we removed any SNP not present in the SNP tables provided by freeze 2 of the DGRP [44] (http://www.hgsc.bcm.tmc.edu/projects/dgrp/). Of the 1,500,000 SNPs initially identified, ~500,000 SNPs remained after applying these filters (Supplemental table 2). SNPs were annotated using SNPeff version 2.0.5 [103]. Short intron annotations were taken from [42].

*Fst estimates*. To estimate average differentiation between populations or between samples collected trough time, we calculated genome-wide average (mean) $F_{ST}$ between pairs of populations. $F_{ST}$ was calculated as,

$$F_{ST} = (H_{total} - H_{with}) / H_{total},$$

where $H_{total}$ is the expected heterozygosity between two populations under panmixia and $H_{with}$ is the heterozygosity averaged between the two populations. Estimates of heterozygosity were corrected for read depth and number of sampled chromosomes by the factor,

$$N_{eff} / (N_{eff} - 1)$$

where,

$$N_{eff} = (N_{chr} * N_{rd} - 1) / (N_{chr} + N_{rd})$$

and where $N_{chr}$ is the number of sampled chromosomes and $N_{rd}$ is the number of reads at any site [104-106].

We performed a parametric permutation analysis to calculate the expected, genome-wide average $F_{ST}$ between pairs of populations under the null hypothesis of panmixia (spatial) or no allele frequency change through time (temporal) conditional on our experimental sampling design. To do so, we calculated the average allele frequency between any two pairs of populations or samples and randomly generated two estimates of allele frequency conditional on the average allele frequency, the number of reads at that site and the number of chromosomes sampled.

To calculate the probability that observed $F_{ST}$ is greater than expected by chance, we generated 500 block bootstrap samples of ~2300 SNPs, where one SNP was drawn per 50kb interval. The probability that the observed $F_{ST}$ distribution is greater than expected by chance is thus,

$$Pr(Obs\ F_{ST} > Exp\ F_{ST}) = E(E(Obs\ F_{ST,i} > Exp\ F_{ST,i})_j),$$

with standard deviation,

$$SD[Pr(Obs\ F_{ST} > Exp\ F_{ST})] = Sd(E(Obs\ F_{ST,i} > Exp\ F_{ST,i})_j),$$

where *i* refers to the $i^{th}$ SNP from $j^{th}$ block bootstrap sample.

*Identification of seasonally and clinally varying polymorphisms*. To identify clinally varying and seasonally oscillating polymorphisms, we used generalized linear models implemented in R 2.10 [107] with binomial error structure and weights proportional to the number of reads sampled at a site and the number of chromosomes sampled (see above, $N_{eff}$). To identify clinal polymorphisms, we regressed allele frequency at each site (excluding all Pennsylvanian samples) on latitude (Supplemental Table 1) according to the form,

$$y_i = lat + \varepsilon_i,$$

where $y_i$ is the observed allele frequencies of the $i^{th}$ SNP and $\varepsilon_i$ is the binomial error given the number of effective reads (see above) at the $i^{th}$ SNP. To identify seasonally oscillating polymorphisms, we regressed allele frequency for the three sets of spring and fall samples on a binary variable corresponding to *spring* or *fall* according to the form,

$$y_i = season + \varepsilon_i.$$

In addition, we modeled allele frequency change through time using generalized linear mixed models (GLMM) implemented in the *lme4* R package [108] and generalized estimation equations (GEE) implemented in the *geepack* R package [109]. We fit GLMMs with the model,

$$y_i = season + (1|population_j) + \varepsilon_i,$$

where $(1|population_j)$ corresponds to the random effect of population $j$ and $\varepsilon_i$ corresponds to the binomial error. We fit GEEs with the model,

$$y_i = season + population_j + \varepsilon_i,$$

where $population_j$ corresponds to the population level strata and $\varepsilon_i$ corresponds to the binomial error fit with either an autoregressive order one correlation structure. q-q plots (Supplemental Fig. 2) demonstrate that these models (clinal and seasonal) fit the bulk of the data adequately, with the exception of the seasonal GEE model which appears to be exceedingly anti-conservative. The false discovery rate was estimated using the Benjamini & Hochberg procedure [110].

For seasonal SNPs, we estimated the cumulative selection coefficient as,

$$S = log(((1/\hat{p}_{Fall})-1)/((1/\hat{p}_{Spring})-1))),$$

where, $\hat{p}_{Fall}$ is the average allele frequency at seasonal SNPs in the fall and $\hat{p}_{Spring}$ is the average allele frequency at seasonal SNPs in the spring.

*Control polymorphisms and the block bootstrap.* Throughout our analysis, we contrasted seasonal SNPs with control polymorphisms (Figs. 2-6). For these analyses, we identified 500 sets of control polymorphisms matched to each seasonal SNP. For each test described in the results, control polymorphisms were identified based on different sets of characteristics that have been shown, or could plausibly, influence the parameter we sought to investigate. In general, we matched seasonal SNPs to control SNPs by chromosome, recombination rate, and allele frequency in either Pennyslvania, North Carolina, North America, and/or Africa. The choice of which population to match allele frequencies was determined by the specific test. These three parameters (chromosome, recombination rate, allele frequency) correspond with many important evolutionary processes as well as genetic patterns (e.g., [111]) and therefore control SNPs will be matched to seasonal SNPs with respect to long-term evolutionary history, gene-density, background levels of genetic variation. In general, we used as many parameters as possible while still identifying a sufficient number of control SNPs for each test and a full list of the matched characters for each test are listed in Supplemental table 3. For continuous characters, such as allele frequency, we typically rounded values so that a sufficient number of unique control sites could be identified. If no matched control SNPs were identified for a seasonal SNP, that seasonal SNP was removed from subsequent analyses.

In addition, we implemented a block-bootstrap procedure to ameliorate positive dependence of our test-statistics due to linkage disequilbrium between seasonal SNPs. We generated 500 sets of seasonal SNPs where one seasonal SNP was sampled from each 50kb consecutive interval of the genome. This block-bootstrap yielded ~850 SNPs that were spaced approximately every 50Kb.

Estimates of expected values (E) of test statistics [e.g. $\log_2$-odds-ratios (Fig. 2C, 3B-C, 6A), $F_{ST}$ (Fig. 2D), probability (Fig. 4B)] and standard deviations (SD) about those expected values were calculated as,

$$E(TS) = E(\ E(TS)_i\ )_j,$$
$$SD(TS) = SD(\ E(TS)_i\ )_j,$$

where $i$ refers to control bootstrap set $i$ and $j$ refers to block bootstrap set $j$ of any test-statistic, $TS$.

*Power calculations*. To calculate statistical power of our experiment and to estimate the expected number of SNPs that are likely to vary repeatedly between seasons and along the cline we used Monte Carlo simulations based on the observed changes in allele frequency between spring and fall at seasonal SNPs or Maine and Florida at clinal SNPs. We calculated statistical power to detect seasonal SNPs as the probability of rejecting the null hypothesis of no repeatable change in allele frequency between spring and fall over three years given our sampling effort (e.g., number of chromosomes from nature and distribution of read depths in our Pennsylvanian samples) at $\alpha < \sim$1e-5, corresponding to observed seasonal q-value of 0.3, conditional on $S$, the cumulative change in allele frequency between seasons calculated from the logistic function. Similarly, we calculated statistical power to detect clinal SNPs as the probability of rejecting the null hypothesis of no change in allele frequency with latitude given our sampling effort at $\alpha < 0.02$, corresponding to the observed clinal q-value of 0.1, conditional on beta, the slope of the relationship between allele frequency and latitude. The expected number of seasonally (clinally) varying SNPs is then, the number of observed seasonal (clinal) SNPs at a particular value of S (beta) divided by the power to detect a seasonal (clinal) SNP at a selection coefficient S.

*Comparison with D. simulans*. To estimate the extent of trans-specific polymorphism between *D. melanogaster* and *D. simulans*, we used *D. simulans* haplotype data available from the DPGP [112] (http://www.dpgp.org/). First, we remapped raw shot-gun sequences of each *D. simulans* strain (GenBank accessions AASS00000000 - AASW00000000) to the latest release of the *D. simulans* reference genome [113] with *bwa* version 0.5.9-r16 using the *bwa-sw* method.

To convert the genomic coordinate system of the new *D. simulans* genome to the *D. melanogaster* genome, we generated a lift-over file using *lastz* [114] and components of the UCSC genome-browser toolkit [115]. Gap parameters corresponded to those used to generate the lift-over file between the first generation *D. simulans* genome and the *D. melanogaster* genome (http://hgdownload.soe.ucsc.edu/goldenPath/dm3/vsDroSim1/). The lift-over file to translate the coordinate system of the second generation *D. simulans* genome to the *D. melanogaster* version 5 genome is available from the Data Dryad under accession number XXXX.

We calculated average pairwise distance between *D. melanogaster* and *D. simulans* haplotypes at seasonal SNPs that were polymorphic in both species and shared the same two alleles by state. We calculated average pairwise distance at two windows surrounding seasonal SNPs, ±1-250bp. Note, we excluded the focal, seasonal SNP. Pairwise distance calculations were performed using the *ape* [116] package in R.

*Forward genetic simulations.* To simulate genome-wide allele frequency change due to cyclic changes in population size and selection at seasonally adaptive polymorphisms, we used a modified version of the forward genetic simulation software SLiM [72]. Source code for the modified version of SLiM is available upon request. In these simulations, we modeled a 20Mb chromosome with constant recombination rate of 2cM/Mb. For all simulations, we seeded the chromosome with 500 neutral mutations randomly placed along the chromosome all starting at 50% initial allele frequency and in complete linkage equilibrium. The number of loci under selection varied between 0 and 30 and loci under temporally heterogeneous selection were placed equidistantly along the chromosome. Selection coefficients for each selected locus were set to produce adaptive oscillations between 40 and 60% frequency every 2 (simulated 'winter') and 10 (simulated 'summer') generations. Genotypic state was assigned randomly to each simulated diploid genome at each selected locus. Population size varied over the course of each simulation. Populations grew exponentially each 'summer' to a maximum population size of $10^5$ over 10 generations. Population size instantaneously crashed at the start of winter to between 5 and $10^4$ individuals and was held constant for two generations. Simulations were run for 100 generations and $F_{ST}$ was estimated from the last three summer-winter cycles.

*Truncation selection model.* To estimate the upper bound of the number of loci that could plausibly respond to seasonally variable selection, we modeled a simple truncation selection scenario. For these models we calculated the expected number of winter adaptive alleles in the fall and the spring as the sum of average allele frequencies of the winter alleles in our fall and spring samples. If the oscillating alleles segregate independently, the variance in the number of winter alleles at any given time follows a Poisson distribution with mean and variance equal to the expected number of winter alleles. Therefore, the proportion of the population in the selected tail over winter is the probability of sampling the expected number of winter alleles in the spring from a Poisson distribution with mean equal to the number of winter alleles in the fall. To vary the number of independently oscillating polymorphisms in the spring and fall, we sub-sampled the number of oscillating polymorphisms 500 times for a range of values.

*References*
1. Gillespie JH (1973) Polymorphism in Random Environments. Theoretical Population Biology 4: 193-195.

2. Ellner S, Sasaki A (1996) Patterns of genetic polymorphism maintained by fluctuating selection with overlapping generations. Theor Popul Biol 50: 31-65.
3. Korol AB, Kirzhner VM, Ronin YI, Nevo E (1996) Cyclical environmental changes as a factor maintaining genetic polymorphism. 2. Diploid selection for an additive trait. Evolution 50: 1432-1441.
4. Ewing EP (1979) Genetic-Variation in a Heterogeneous Environment. 7. Temporal and Spatial Heterogeneity in Infinite Populations. American Naturalist 114: 197-212.
5. Gillespie JH, Langley CH (1974) General Model to Account for Enzyme Variation in Natural-Populations. Genetics 76: 837-884.
6. Hoekstra RF (1978) Sufficient Conditions for Polymorphism with Cyclical Selection in a Subdivided Population. Genetical Research 31: 67-73.
7. Haldane JBS, Jayakar SD (1962) Polymorphism Due to Selection of Varying Direction. Journal of Genetics 58: 237-242.
8. Hedrick PW (1976) Genetic-Variation in a Heterogeneous Environment. 2. Temporal Heterogeneity and Directional Selection. Genetics 84: 145-157.
9. Gershenson S (1945) Evolutionary Studies on the Distribution and Dynamics of Melanism in the Hamster (*Cricetus cricetus L*). 1. Distribution of Black Hamsters in the Ukrainian and Bashkirian Soviet Socialist Republics (Ussr). Genetics 30: 207-232.
10. Grant PR, Grant BR (2002) Unpredictable evolution in a 30-year study of Darwin's finches. Science 296: 707-711.
11. Tarwater CE, Beissinger SR (2013) Opposing selection and environmental variation modify optimal timing of breeding. Proc Natl Acad Sci U S A 110: 15365-15370.
12. Brown CR, Brown MB, Roche EA (2013) Fluctuating viability selection on morphology of cliff swallows is driven by climate. J Evol Biol 26: 1129-1142.
13. Wall S, Carter MA, Clarke B (1980) Temporal Changes of Gene-Frequencies in *Cepaea hortensis*. Biological Journal of the Linnean Society 14: 303-317.
14. Brakefield PM (1985) Differential Winter Mortality and Seasonal Selection in the Polymorphic Ladybird *Adalia bipunctata* (L) in the Netherlands. Biological Journal of the Linnean Society 24: 189-206.
15. Hairston NG, Dillon TA (1990) Fluctuating Selection and Response in a Population of Fresh-Water Copepods. Evolution 44: 1796-1805.
16. Bradshaw WE (1973) Homeostasis and Polymorphism in Vernal Development of *Chaoborus americanus*. Ecology 54: 1247-1259.
17. Vavrek MC, McGraw JB, Yang HS (1996) Within-population variation in demography of *Taraxacum officinale*: Maintenance of genetic diversity. Ecology 77: 2098-2107.
18. Axenovich TI, Zorkoltseva IV, Akberdin IR, Beketov SV, Kashtanov SN, et al. (2007) Inheritance of litter size at birth in farmed arctic foxes (*Alopex lagopus*, Canidae, Carnivora). Heredity 98: 99-105.
19. Vorburger C (2006) Temporal dynamics of genotypic diversity reveal strong clonal selection in the aphid *Myzus persicae*. Journal of Evolutionary Biology 19: 97-107.
20. Niklasson M, Tomiuk J, Parker ED (2004) Maintenance of clonal diversity in *Dipsa bifurcata* (Fallen, 1810) (Diptera : Lonchopteridae). I. Fluctuating seasonal selection moulds long-term coexistence. Heredity 93: 62-71.

21. Tomiuk J, Niklasson M, Parker ED (2004) Maintenance of clonal diversity in *Dipsa bifurcata* (Fallen, 1810) (Diptera : Lonchopteridae). II. Diapause stabilizes clonal coexistence. Heredity 93: 72-77.
22. Templeton A, Johnston JS (1982) Life history evolution under pleiotropy and K-selection in a natural population of *Drosophila mercatorum*. In: Barker JSF, Starmer WT, editors. Ecological Genetics and Evolution: The Cactus-Yeast-Drosophila Model System. New York: Academic Press. pp. 225-239.
23. Hart-Schmidt RA (2012) Geographically patterned variation in diapause and its relationship to other climate-associated phenotypes and genotypes of *Drosophila americana*: University of Iowa.
24. Rodriguez-Trelles F, Tarrio R, Santos M (2013) Genome-wide evolutionary response to a heat wave in Drosophila. Biol Lett 9: 20130228.
25. Bell G (2010) Fluctuating selection: the perpetual renewal of adaptation in variable environments. Philos Trans R Soc Lond B Biol Sci 365: 87-97.
26. Siepielski AM, DiBattista JD, Carlson SM (2009) It's about time: the temporal dynamics of phenotypic selection in the wild. Ecol Lett 12: 1261-1276.
27. Dobzhansky T (1943) Genetics of Natural Populations IX. Temporal Changes in the Composition of Populations of *Drosophila pseudoobscura*. Genetics 28: 162-186.
28. Dobzhansky T, Ayala FJ (1973) Temporal Frequency Changes of Enzyme and Chromosomal Polymorphisms in Natural Populations of Drosophila. Proceedings of the National Academy of Sciences of the United States of America 70: 680-683.
29. Mueller LD, Barr LG, Ayala FJ (1985) Natural-Selection Vs Random Drift - Evidence from Temporal Variation in Allele Frequencies in Nature. Genetics 111: 517-554.
30. Nielsen KM, Hoffmann AA, Mckechnie SW (1985) Population-Genetics of the Metabolically Related Adh, Gpdh and Tpi Polymorphisms in *Drosophila melanogaster*. 2. Temporal and Spatial Variation in an Orchard Population. Genetics Selection Evolution 17: 41-58.
31. Knibb WR (1986) Temporal Variation of *Drosophila melanogaster* Adh Allele Frequencies, Inversion Frequencies, and Population Sizes. Genetica 71: 175-190.
32. Rodriguez-Trelles F, Alvarez G, Zapata C (1996) Time-series analysis of seasonal changes of the O inversion polymorphism of *Drosophila subobscura*. Genetics 142: 179-187.
33. Ananina G, Peixoto AA, Bitner-Mathe BC, Souza WN, da Silva LB, et al. (2004) Chromosomal inversion polymorphism in *Drosophila mediopunctata*: seasonal, altitudinal, and latitudinal variation. Genetics and Molecular Biology 27: 61-69.
34. Bouletreaumerle J, Fouillet P, Terrier O (1992) Clinal and Seasonal-Variations in Initial Retention Capacity of Virgin *Drosophila melanogaster* Females as a Strategy for Fitness. Evolutionary Ecology 6: 223-242.
35. Schmidt PS, Conde DR (2006) Environmental heterogeneity and the maintenance of genetic variation for reproductive diapause in *Drosophila melanogaster*. Evolution 60: 1602-1611.
36. Stalker HD, Carson HL (1949) Seasonal Variation in the Morphology of *Drosophila robusta* Sturtevant. Evolution 3: 330-343.

37. Tantawy AO (1964) Studies on Natural-Populations of Drosophila. 3. Morphological and Genetic-Differences of Wing Length in *Drosophila melanogaster* and *Drosophila simulans* in Relation to Season. Evolution 18: 560-570.
38. Miyo T, Akai S, Oguma Y (2000) Seasonal fluctuation in susceptibility to insecticides within natural populations of *Drosophila melanogaster*: empirical observations of fitness costs of insecticide resistance. Genes & Genetic Systems 75: 97-104.
39. Dev K, Chahal J, Parkash R, Kataria SK (2013) Correlated Changes in Body Melanisation and Mating Traits of *Drosophila melanogaster*: A Seasonal Analysis. Evolutionary Biology 40: 366-376.
40. Zhu Y, Bergland AO, Gonzalez J, Petrov DA (2012) Empirical Validation of Pooled Whole Genome Population Re-Sequencing in Drosophila melanogaster. Plos One 7: e41901.
41. Wright S (1943) Isolation by Distance. Genetics 28: 114-138.
42. Lawrie DS, Messer PW, Hershberg R, Petrov DA (2013) Strong Purifying Selection at Synonymous Sites in *Drosophila melanogaster*. PloS Genetics 9(5): e1003527.
43. Pechmann S, Frydman J (2013) Evolutionary conservation of codon optimality reveals hidden signatures of cotranslational folding. Nat Struct Mol Biol 20: 237-243.
44. Mackay TF, Richards S, Stone EA, Barbadilla A, Ayroles JF, et al. (2012) The *Drosophila melanogaster* Genetic Reference Panel. Nature 482: 173-178.
45. Kapun M, van Schalkwyk H, McAllister B, Flatt T, Schlotterer C (2013) Inference of chromosomal inversion dynamics from Pool-Seq data in natural and laboratory populations of *Drosophila melanogaster*. arXiv 1307.2461.
46. Corbett-Detig RB, Hartl DL (2012) Population genomics of inversion polymorphisms in *Drosophila melanogaster*. PLoS Genet 8: e1003056.
47. Gibert P, Moreteau B, Petavy G, Karan D, David JR (2001) Chill-coma tolerance, a major climatic adaptation among Drosophila species. Evolution 55: 1063-1068.
48. Karan D, Dahiya N, Munjal AK, Gibert P, Moreteau B, et al. (1998) Desiccation and starvation tolerance of adult Drosophila: Opposite latitudinal clines in natural populations of three different species. Evolution 52: 825-831.
49. Kenny MC, Wilton A, Ballard JWO (2008) Seasonal trade-off between starvation resistance and cold resistance in temperate wild-caught *Drosophila simulans*. Australian Journal of Entomology 47: 20-23.
50. Parkash R, Sharma V, Kalra B (2009) Impact of body melanisation on desiccation resistance in montane populations of *D. melanogaster*: Analysis of seasonal variation. Journal of Insect Physiology 55: 898-908.
51. Schmidt PS, Paaby AB, Heschel MS (2005) Genetic variance for diapause expression and associated life histories in *Drosophila melanogaster*. Evolution 59: 2616-2625.
52. Sisodia S, Singh BN (2010) Resistance to environmental stress in *Drosophila ananassae*: latitudinal variation and adaptation among populations. Journal of Evolutionary Biology 23: 1979-1988.
53. Huang W, Richards S, Carbone MA, Zhu DH, Anholt RRH, et al. (2012) Epistasis dominates the genetic architecture of Drosophila quantitative traits. Proceedings of the National Academy of Sciences of the United States of America 109: 15553-15559.
54. Pool JE, Corbett-Detig RB, Sugino RP, Stevens KA, Cardeno CM, et al. (2012) Population Genomics of Sub-Saharan *Drosophila melanogaster*: African Diversity and Non-African Admixture. PLoS Genet 8: e1003080.
55. Klein J, Sato A, Nagl S, O'hUigin C (1998) Molecular trans-species polymorphism. Annual Review of Ecology and Systematics 29: 1-+.
56. Leffler EM, Gao Z, Pfeifer S, Segurel L, Auton A, et al. (2013) Multiple instances of ancient balancing selection shared between humans and chimpanzees. Science 339: 1578-1582.
57. Welbergen P, van Dijken FR, Scharloo W, Kohler W (1992) The genetic basis of sexual isolation between *Drosophila melanogaster* and *D. simulans*. Evolution 46: 1385-1398.
58. Sturtevant AH (1920) Genetic studies on *Drosophila simulans*. I. Introduction. Hybrids with *Drosophila melanogaster*. Genetics 5: 488-500.
59. Tamura K, Subramanian S, Kumar S (2004) Temporal patterns of fruit fly (Drosophila) evolution revealed by mutation clocks. Mol Biol Evol 21: 36-44.
60. Gao Z, Przeworski M, Sella G (2014) Footprints of ancient balanced polymorphisms in genetic vartiation data. arXiv: 1401.7589.
61. Karasov T, Messer PW, Petrov DA (2010) Evidence that adaptation in Drosophila is not limited by mutation at single sites. PLoS Genet 6: e1000924.
62. Mckenzie JA (1980) An Ecological Study of the Alcohol-Dehydrogenase (Adh) Polymorphism of *Drosophila melanogaster*. Australian Journal of Zoology 28: 709-716.
63. Mcinnis DO, Schaffer HE, Mettler LE (1982) Field Dispersal and Population Sizes of Native Drosophila from North-Carolina. American Naturalist 119: 319-330.
64. Powell J (1997) Progress and prospects in evolutionary biology: The Drosophila Model. New York: Oxford University Press.
65. Gravot E, Huet M, Veuille M (2004) Effect of breeding structure on population genetic parameters in Drosophila. Genetics 166: 779-788.
66. Bastide H, Betancourt A, Nolte V, Tobler R, Stobe P, et al. (2013) A Genome-Wide, Fine-Scale Map of Natural Pigmentation Variation in *Drosophila melanogaster*. Plos Genetics 9.
67. Kirkpatrick M (2000) Reinforcement and divergence under assortative mating. Proceedings of the Royal Society B-Biological Sciences 267: 1649-1655.
68. Lewontin RC (1964) The Interaction of Selection and Linkage. I. General Considerations; Heterotic Models. Genetics 49: 49-67.
69. Giesel JT (1977) A model of functional epistasis and linkage disequilibrium in populations with overlapping generations. Genetics 86: 679-686.
70. Band HT, Ives PT (1963) Genetic-Structure of Populations .1. Nature of Genetic Load in South Amherst Population of *Drosophila melanogaster*. Evolution 17: 198-&.
71. Ives PT, Band HT (1986) Continuing Studies on the South Amherst *Drosophila melanogaster* Natural-Population during the 1970s and 1980s. Evolution 40: 1289-1302.

72. Messer PW (2013) SLiM: Simulating Evolution with Selection and Linkage. Genetics 194: 1037-+.
73. Comeron JM, Ratnappan R, Bailin S (2012) The Many Landscapes of Recombination in *Drosophila melanogaster*. Plos Genetics 8.
74. Schmidt PS, Conde DR (2006) Environmental heterogeneity and the maintenance of genetic variation for reproductive diapause in *Drosophila melanogaster*. Evolution 60: 1602-1611.
75. Ives PT (1970) Futher genetic studies on the South Amherst population of *Drosophila melanogaster*. Evolution 24: 507-518.
76. Coyne JA, Milstead B (1987) Long-Distance Migration of Drosophila. 3. Dispersal of *Drosophila melanogaster* Alleles from a Maryland Orchard. American Naturalist 130: 70-82.
77. Umina PA (2005) A rapid shift in a classic clinal pattern in Drosophila reflecting climate change. Nature 308: 691-693.
78. Berry A, Kreitman M (1993) Molecular analysis of an allozyme cline: alcohol dehydrogenase in *Drosophila melanogaster* on the east coast of North America. Genetics 134: 869-893.
79. Lewontin RC (1965) Selection for colonizing ability. In: Baker HG, Stebbins GL, editors. The Genetics of Colonizing Species First International Union of Biological Sciences Symposia on General Biology, Aslimoar, Calif, 1964. New York: Academic Press. pp. 77-94.
80. Schmidt PS, Paaby AB (2008) Reproductive diapause and life-history clines in North American populations of *Drosophila melanogaster*. Evolution 62: 1204-1215.
81. Schmidt PS, Matzkin L, Ippolito M, Eanes WF (2005) Geographic variation in diapause incidence, life-history traits, and climatic adaptation in *Drosophila melanogaster*. Evolution 59: 1721-1732.
82. Arthur AL, Weeks AR, Sgro CM (2008) Investigating latitudinal clines for life history and stress resistance traits in *Drosophila simulans* from eastern Australia. Journal of Evolutionary Biology 21: 1470-1479.
83. James AC, Azevedo RB, Partridge L (1997) Genetic and environmental responses to temperature of *Drosophila melanogaster* from a latitudinal cline. Genetics 146: 881-890.
84. Robinson SJ, Zwaan B, Partridge L (2000) Starvation resistance and adult body composition in a latitudinal cline of *Drosophila melanogaster*. Evolution 54: 1819-1824.
85. Lazzaro BP, Flores HA, Lorigan JG, Yourth CP (2008) Genotype-by-environment interactions and adaptation to local temperature affect immunity and fecundity in *Drosophila melanogaster*. Plos Pathogens 4.
86. Singh RS, Long AD (1992) Geographic variation in Drosophila: From molecules to morphology and back. Trends Ecol Evol 7: 340-345.
87. Ayrinhac A, Debat V, Gibert P, Kister AG, Legout H, et al. (2004) Cold adaptation in geographical populations of *Drosophila melanogaster*: phenotypic plasticity is more important than genetic variability. Functional Ecology 18: 700-706.
88. Hoffmann AA, Anderson A, Hallas R (2002) Opposing clines for high and low temperature resistance in *Drosophila melanogaster*. Ecology Letters 5: 614-618.

89. Overgaard J, Kristensen TN, Mitchell KA, Hoffmann AA (2011) Thermal Tolerance in Widespread and Tropical Drosophila Species: Does Phenotypic Plasticity Increase with Latitude? American Naturalist 178: S80-S96.
90. Hoffmann AA, Harshman LG (1999) Desiccation and starvation resistance in Drosophila: patterns of variation at the species, population and intrapopulation levels. Heredity 83: 637-643.
91. Gilchrist GW, Jeffers LM, West B, FOlk DG, Suess J, et al. (2008) Clinal patterns of desiccation and starvation resistance in ancestral and invading populations of *Drosophila subobscura*. Evolutionary Applications 1: 513-523.
92. Ayroles JF, Carbone MA, Stone EA, Jordan KW, Lyman RF, et al. (2009) Systems genetics of complex traits in *Drosophila melanogaster*. Nat Genet 41: 299-307.
93. Reynolds LA (2013) The effects of starvation selection on *Drosophila melanogaster* life history and development: UNLV.
94. David JR, Capy P (1988) Genetic-Variation of *Drosophila melanogaster* Natural-Populations. Trends in Genetics 4: 106-111.
95. Sezgin E, Duvernell DD, Matzkin LM, Duan Y, Zhu CT, et al. (2004) Single-locus latitudinal clines and their relationship to temperate adaptation in metabolic genes and derived alleles in *Drosophila melanogaster*. Genetics 168: 923-931.
96. Markow TA, O'Grady PM (2006) Drosophila: A guide to species identification and use. London: Academic Press.
97. Langley CH, Stevens K, Cardeno C, Lee YC, Schrider DR, et al. (2012) Genomic variation in natural populations of *Drosophila melanogaster*. Genetics 192: 533-598.
98. Ginzburg L, Colyvan M (2003) Ecological orbits: How planets move and populations grow. New York: Oxford University Press.
99. Li H, Durbin R (2009) Fast and accurate short read alignment with Burrows-Wheeler transform. Bioinformatics 25: 1754-1760.
100. Li H, Handsaker B, Wysoker A, Fennell T, Ruan J, et al. (2009) The Sequence Alignment/Map format and SAMtools. Bioinformatics 25: 2078-2079.
101. McKenna A, Hanna M, Banks E, Sivachenko A, Cibulskis K, et al. (2010) The Genome Analysis Toolkit: a MapReduce framework for analyzing next-generation DNA sequencing data. Genome Res 20: 1297-1303.
102. Bansal V (2010) A statistical method for the detection of variants from next-generation resequencing of DNA pools. Bioinformatics 26: i318-324.
103. Cingolani P, Platts A, Wang le L, Coon M, Nguyen T, et al. (2012) A program for annotating and predicting the effects of single nucleotide polymorphisms, SnpEff: SNPs in the genome of *Drosophila melanogaster* strain w1118; iso-2; iso-3. Fly (Austin) 6: 80-92.
104. Nei M (1987) Molecular Evolutionary Genetics. New York: Columbia University Press. 512 p.
105. Kolaczkowski B, Kern AD, Holloway AK, Begun DJ (2010) Genomic Differentiation Between Temperate and Tropical Australian Populations of *Drosophila melanogaster*. Genetics 187: 245-260.
106. Feder AF, Petrov DA, Bergland AO (2012) LDx: Estimation of Linkage Disequilibrium from High-Throughput Pooled Resequencing Data. Plos One 7.

107. R Core Development Team (2009) R: A language and environment for statistical computing. Vienna, Austria: R Foundation for Statistical Computing.
108. Bates D, Maechler M, Bolker B (2011) lme4: Linear mixed-effects models using S4 classes. R package version 0.999375-42. http://CRAN.R-project.org/package=lme4
109. Højsgaard S, Halekoh U, Yan J (2006) The R Package geepack for Generalized Estimating Equations. Journal of Statistical Software, 15: 1-11.
110. Benjamini Y, Hochberg Y (1995) Controlling the false discovery rate: a practical and powerful approach ot multiple testing. Journal of the Royal Statistical Society B 57: 289-300.
111. Begun DJ, Aquadro CF (1992) Levels of naturally occurring DNA polymorphism correlate with recombinatin rates in *D. melanogaster*. Nature 356: 519-520.
112. Begun DJ, Holloway AK, Stevens K, Hillier LW, Poh YP, et al. (2007) Population genomics: whole-genome analysis of polymorphism and divergence in *Drosophila simulans*. PLoS Biol 5: e310.
113. Hu TT, Eisen MB, Thornton KR, Andolfatto P (2013) A second-generation assembly of the *Drosophila simulans* genome provides new insights into patterns of lineage-specific divergence. Genome Research 23: 89-98.
114. Harris RS (2007) Improved pairwise alignment of genomic DNA: The Pennsylvania State University.
115. Kent WJ, Sugnet CW, Furey TS, Roskin KM, Pringle TH, et al. (2002) The human genome browser at UCSC. Genome Research 12: 996-1006.
116. Paradis E, Claude J, Strimmer K (2004) APE: analyses of phylogenetics and evolution in R language. Bioinformatics 20: 289-290.

*Acknowledgements*. This work was supported by NIH NRSA fellowship F32GM097837 to AOB, by NSF GRF DGE-0822 to ELB, by NIH RO1GM100366 grant to PSS and DAP, by NIH RO1GM097415 grant to DAP and by NSF DEB 0921307 to PSS. AOB, ELB, KRO, PSS, DAP conceived of the experiment. AOB, ELB, KRO and PSS collected the flies used in this study. AOB, PSS and DAP performed analyses. AOB, PSS and DAP wrote the paper. Raw short read data have been deposited to NCBI Short Read Archive under accession numbers XXXX. Annotated allele frequency estimates, in Variant Call Format, have been deposited to Data Dryad under accession number XXXX. *D. simulans* lift-over files have been deposited to Data Dryad under accession number XXXX.

*Figure legends*

*Figure 1. Experimental design and genomic turnover through time and space.* (A) Map of sampling locations in North America used in this study. Grey boxes represent individual samples from each locale. Genome-wide differentiation among spatially (B) and temporally (C) separated samples, measured as genome-wide average $F_{ST}$. Note: Pennsylvanian samples are not represented in (B) and the negative $F_{ST}$ in (B) results from the conservative correction of heterozygosity [102,103]. Error bars represent 95% confidence intervals based on 500 blocked bootstrap samples of ~2000 SNPs.

*Figure 2. Genomic features of seasonal SNPs.* (A) Allele frequency change at each of the ~1750 seasonal SNPs. Allele frequencies are polarized so that spring allele frequencies

are higher than fall allele frequencies. (B) Power to detect seasonal SNPs (black line) is limited and we estimate that we have only identified ~10% (red line) of all SNPs that repeatedly change in frequency through time (black line). (C) $\log_2$ odds ratio that seasonal SNPs are annotated for each class of genetic element relative to control polymorphisms. (D) Seasonal $F_{ST}$ surrounding seasonal SNPs decays to background levels by ~500bp. (E) Allele frequency estimates at seasonal SNPs outside any large, cosmopolitan inversion (non-inv) or within the cosmopolitian inversions (diamonds) during the spring (blue) or fall (red). Allele frequency estimates at SNPs perfectly linked to the inversion during the spring and fall are denoted by circles. Error bars (C) and confidence bands (D) represent 95% confidence intervals based on blocked bootstrap resampling.

*Figure 3*. *Spatial and temporal variation in allele frequencies*. (A) Genomic distribution of clinal (black line) and seasonal SNPs (red line) per megabase per common polymorphism used in this study (Supplemental Table 1). (B). $\log_2$ odds ratio that SNPs with spatial $F_{ST}$ greater than or equal to value on *x*-axis are enriched for seasonal SNPs versus control SNPs. (C) $\log_2$ odds ratio that SNPs with $-\log_{10}$(spatial q-value) greater than or equal to value on *x*-axis are enriched for seasonal SNPs versus control SNPs. (D) Absolute difference between average spring (blue) and fall (red) frequencies in the Pennsylvanian population and frequency estimates along the cline. Confidence bands represent 95% confidence intervals based on blocked bootstrap resampling.

*Figure 4*. *Adaptive evolution to frost*. (A) Temperature records at a weather station close to the focal orchard. Grey lines indicate collection dates for pre- and post-frost samples. (B) Probability that post-frost allele frequencies at seasonal and control SNPs overshoot the long-term average (based on 2009 and 2010 estimates) allele frequency at each site. Confidence intervals based on blocked bootstrap resampling.

*Figure 5*. *Association with seasonally variable phenotypes*. (A) $\log_2$ odds ratio that the winter favored allele is associated with fast chill coma recovery for SNPs associated with chill coma recovery with GWAS $-\log_{10}$(p) greater than or equal to value on x-axis for seasonal SNPs versus control SNPs. (B) $\log_2$ odds ratio that the winter favored allele is associated with increased starvation tolerance for SNPs associated with starvation tolerance with GWAS $-\log_{10}$(p) greater than or equal to value on x-axis for seasonal SNPs versus control SNPs. Error bars represent 95% confidence intervals based on blocked bootstrap resampling.

*Figure 6*. *Long term balancing selection*. $\log_2$(odds ratio) that seasonal SNPs are polymorphic and identical by state among 6 lineages of *D. simulans* relative to control SNPs. Error bars represent 95% confidence intervals based on blocked bootstrap resampling.

*Figure 7*. *Demographic models*. (A) Expected value of $F_{ST}$ between simulated spring and fall samples (y-axis), conditional on overwintering effective population size and the number of seasonally adaptive alleles (color key). Dotted line represents observed average, genome-wide after $F_{ST}$ between spring and fall samples from the Pennsylvanian

population. (B) Expected number of SNPs that would vary repeatedly between seasons three times in a row conditional on founding deme size for a simple model of recolonization of the orchard population. Dotted line represents the observed number of seasonal SNPs and the corresponding founding deme size required, in this case 5 flies. (C) Minimum population size (y-axis) for the required for varying number of seasonally selected loci (x-axis) under a truncation selection model assuming independent response to selection at each locus. Dotted line represents our best guess of fall population size and corresponding number of loci that could independently respond to truncation selection. Confidence bands based on resampling of observed allele frequency change at seasonal SNPs.

*Supplemental Figure 1. Genomic turnover through space and time – average $F_{ST}$.* Probability that genome-wide average $F_{ST}$ among populations sampled along the cline (A) and through time (B) is greater than expected by chance conditional on our sampling design and panmixia among spatially separated populations or no allele frequency change through time, respectively. Points represent mean $F_{ST}$, error bars represent 95% confidence intervals based on blocked-bootstrap resampling.

*Supplemental Figure 2. q-q plots and congruence of GLM, GLMM and GEE models.* (A-C) Standard q-q plots of *p-values* of GLM, GLMM and GEE models, respectively. q-q plots show that GLM and GLMM models fit the bulk of the genome well whereas GEE models appear to be anti-conservative. (D) $\log_2$(odds-ratio) that the top 1750 seasonal SNPs identified with the GLM model are among the top 1750 seasonal SNPs identified with the GLMM model. (E) $\log_2$(odds-ratio) that the top 1750 seasonal SNPs identified with the GLM model are among the top 1750 seasonal SNPs identified with the GEE model.

*Supplemental Figure 3. Genomic turnover through time excluding SNPs within 1Kb of seasonal SNPs.* (A) Genome-wide average $F_{ST}$ between samples of flies collected through time, excluding SNPs within 1Kb of seasonal SNPs. (B) Probability that genome-wide $F_{ST}$ between pairs of samples collected through time is greater than expected by chance given the null hypothesis of no allele frequency change through time and our sampling design. Solid line represents predicted relationship between genome-wide $F_{ST}$ and time excluding SNPs within 1Kb; dashed line represents predicted relationship between genome-wide $F_{ST}$ for all common SNPs and time. The similarity between the solid and dashed line demonstrates that SNPs near seasonal SNPs are not driving genome-wide patterns of $F_{ST}$ through time. Points represent mean $F_{ST}$, error bars represent 95% confidence intervals based on blocked-bootstrap resampling.

*Supplemental Figure 4. Enrichment among cosmopolitan inversions.* $\log_2$ odds ratio that seasonal SNPs are enriched among the large cosmopolitan inversions relative to control polymorphisms. Inversion breakpoints are defined as ±2.5Mb from the proximal or distal breakpoints. Error bars represent 95% confidence intervals based on blocked bootstrap resampling.

*Supplemental Figure 5*. *Spatial $F_{ST}$ and clinal* q-*value*. Scatter plot of the relationship between spatial $F_{ST}$ (x-axis) and $-\log_{10}$(clinal *q*-value). Colors of the hexagons represent the density of points in that interval.

*Supplemental Figure 6*. *Power to detect clinal SNPs*. Power to detect clinal SNPs (black line) is moderate and we estimate that we have identified ~50% (red line) of all SNPs that change in frequency monotonically with latitude (black line).

*Supplemental Figure 7*. *Site frequency spectrum of seasonal samples*. Unfolded site frequency spectrum of spring (blue) and fall (red) samples from 2009-2010 (A) and 2010-2011 (B). Solid lines represent observed site frequency spectra, dashed lines represent simulated spring site frequency spectra conditional on one generation of bottleneck to 20 individuals and dotted lines represent simulated spring site frequency spectra conditional on two generations of bottleneck to 20 individuals. The increase in low frequency alleles in the spring 2010 sample (B, blue line) is due to the high coverage of this library. Site frequency spectra only included SNPs with allele frequencies greater than 2/(read depth) or less than 1 – 2/(read depth) to account for sequencing errors.

*Supplemental Table 1*. Population sampling locales.

*Supplemental Table 2*. Basic SNP statistics.

*Supplemental Table 3*. Table of control characteristics.

**Figure 1.**

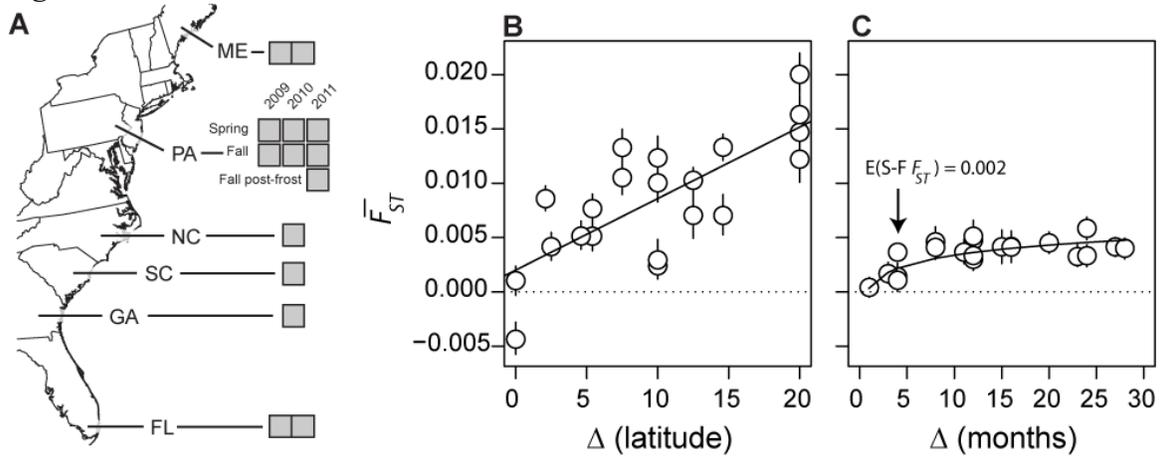

**Figure 2.**

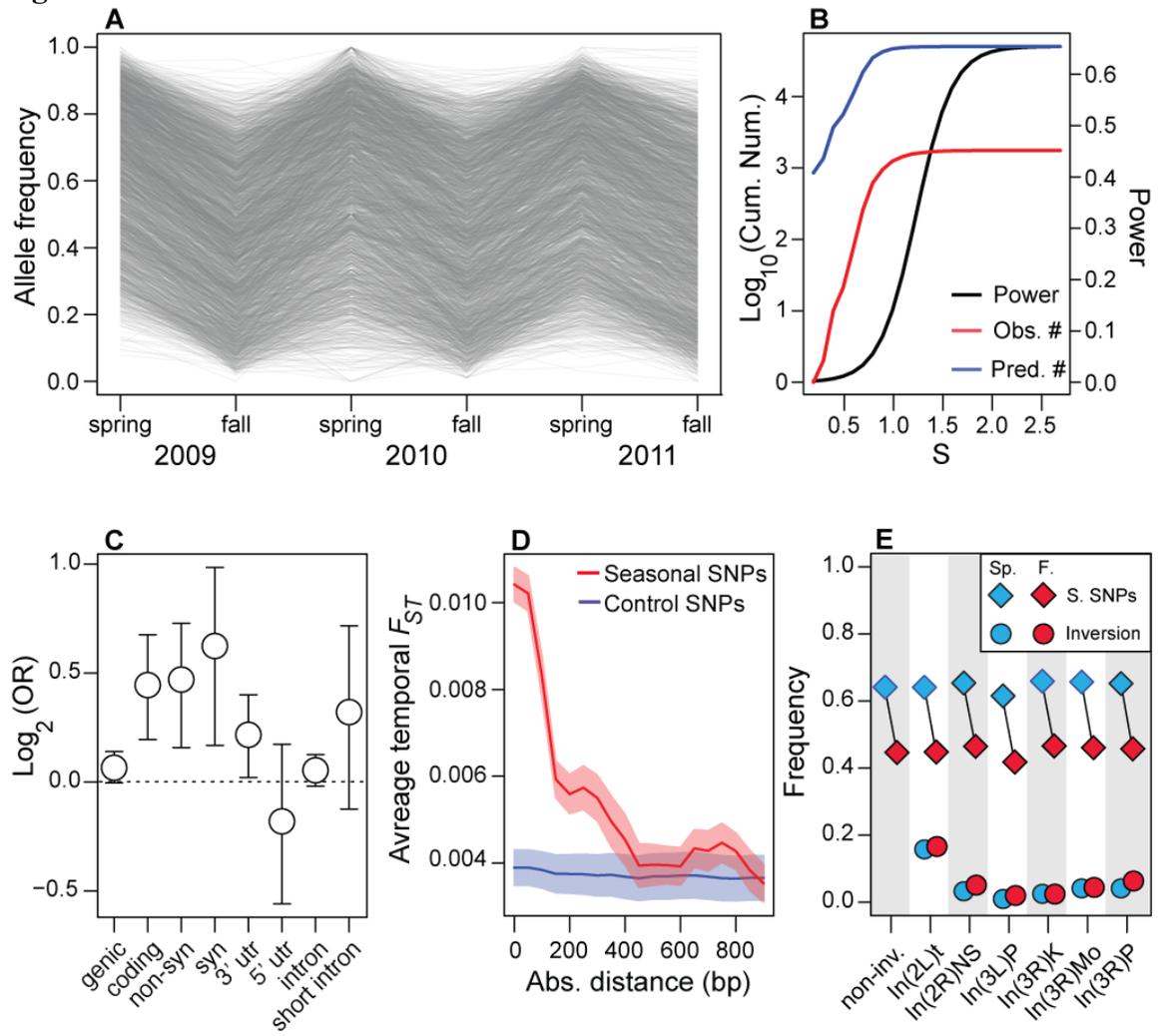

**Figure 3.**

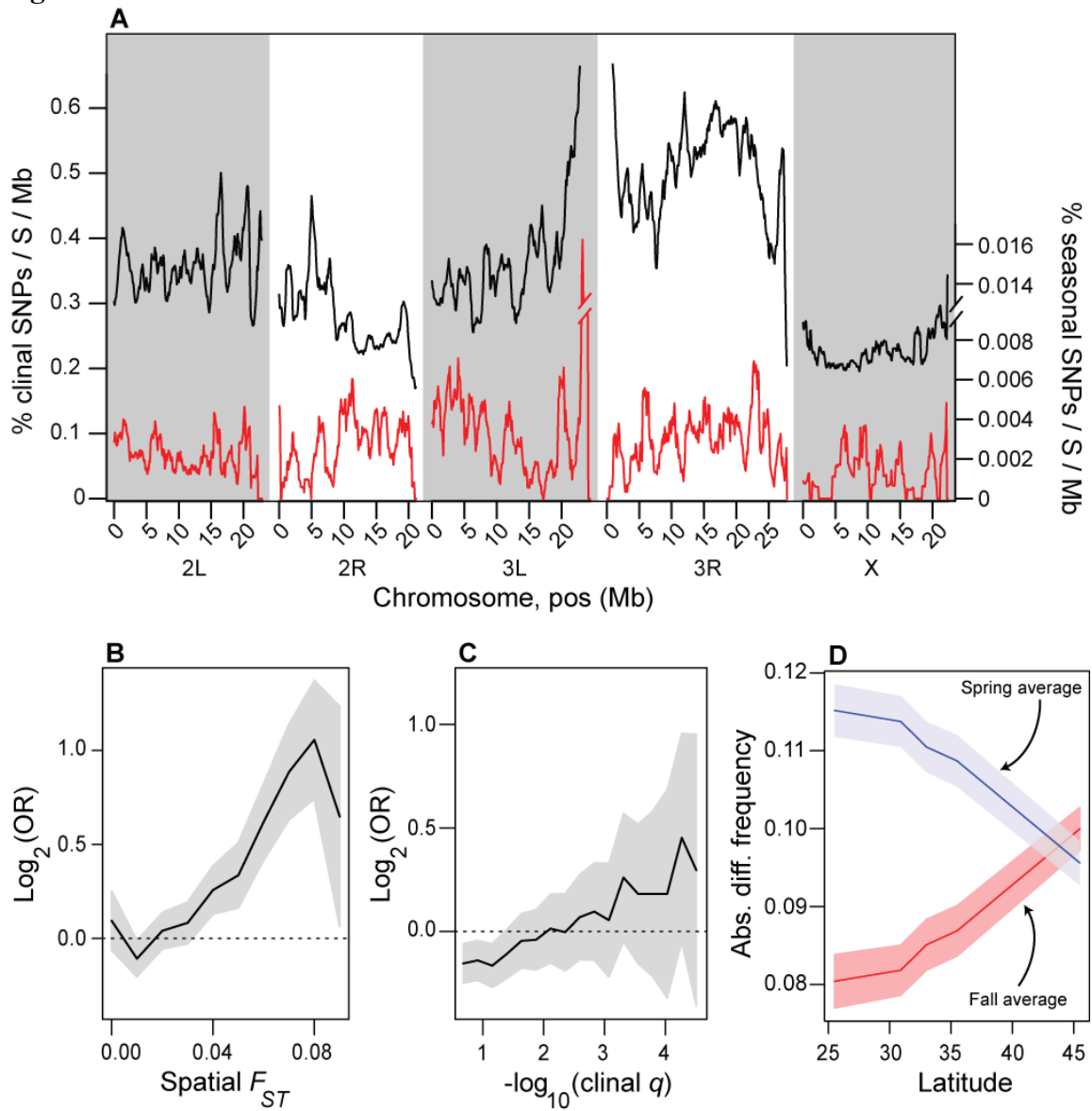

**Figure 4.**

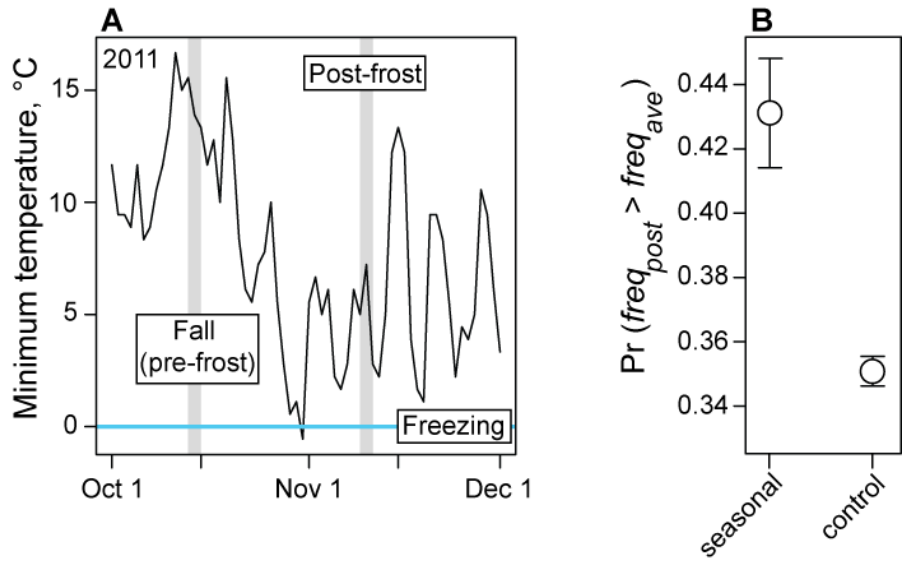

**Figure 5.**

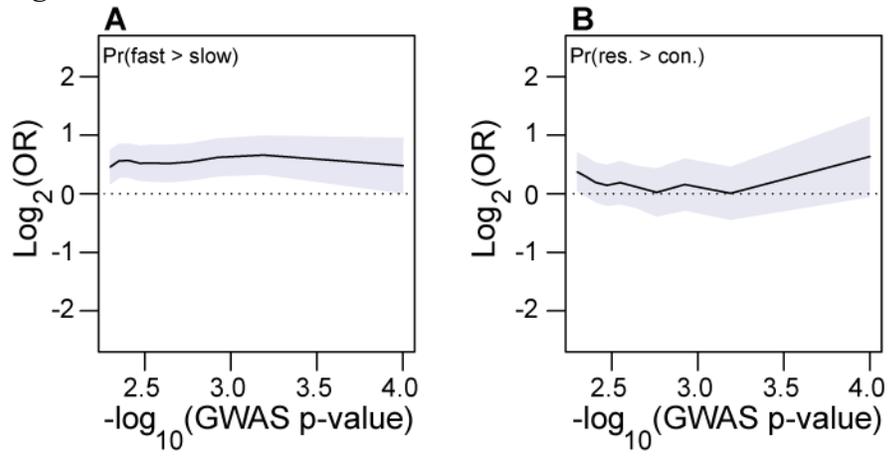

**Figure 6.**

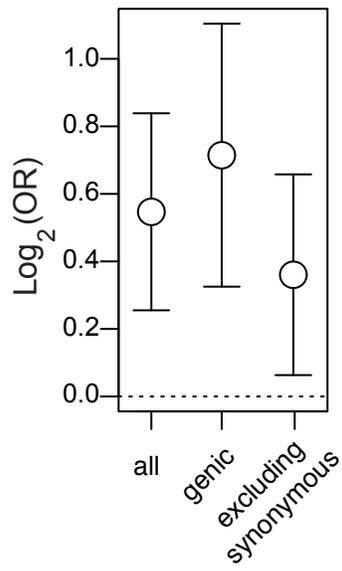

**Figure 7.**

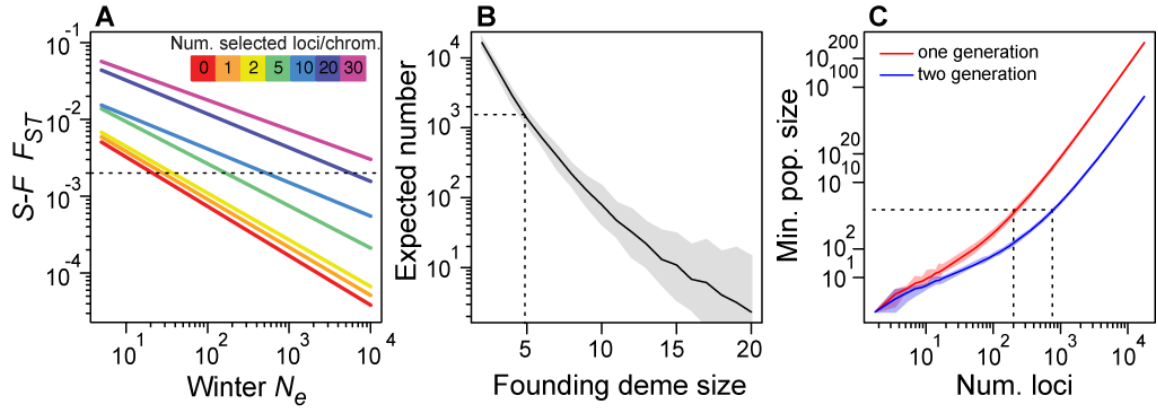

**Supplemental Figure 1.**

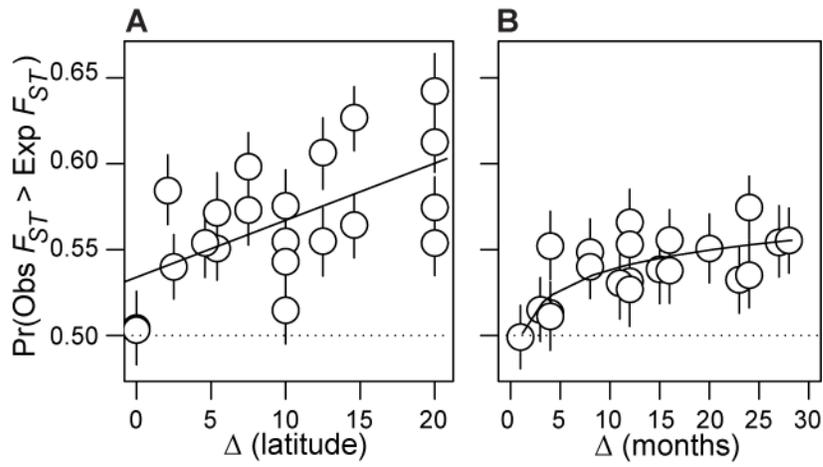

**Supplemental Figure 2.**

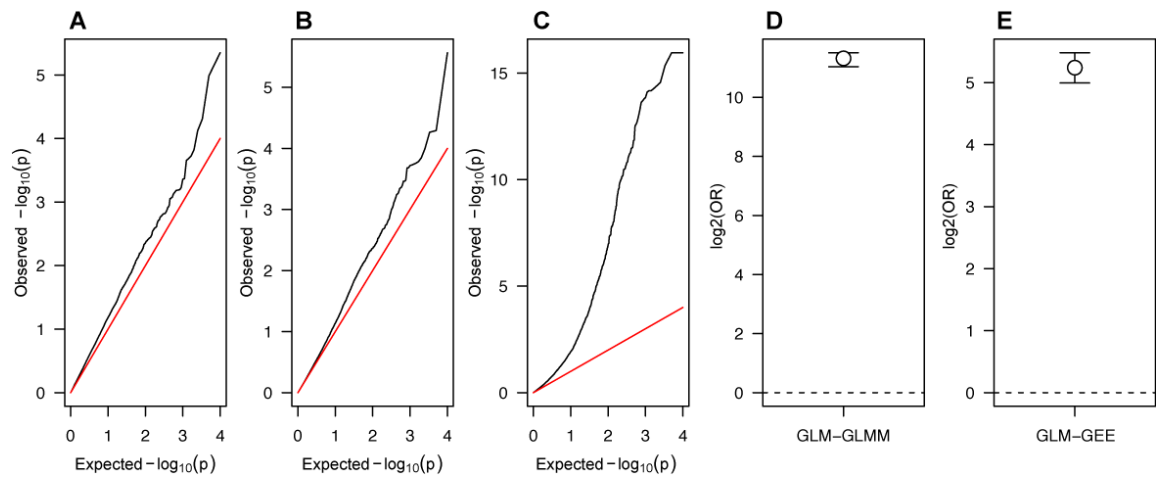

**Supplemental Figure 3.**

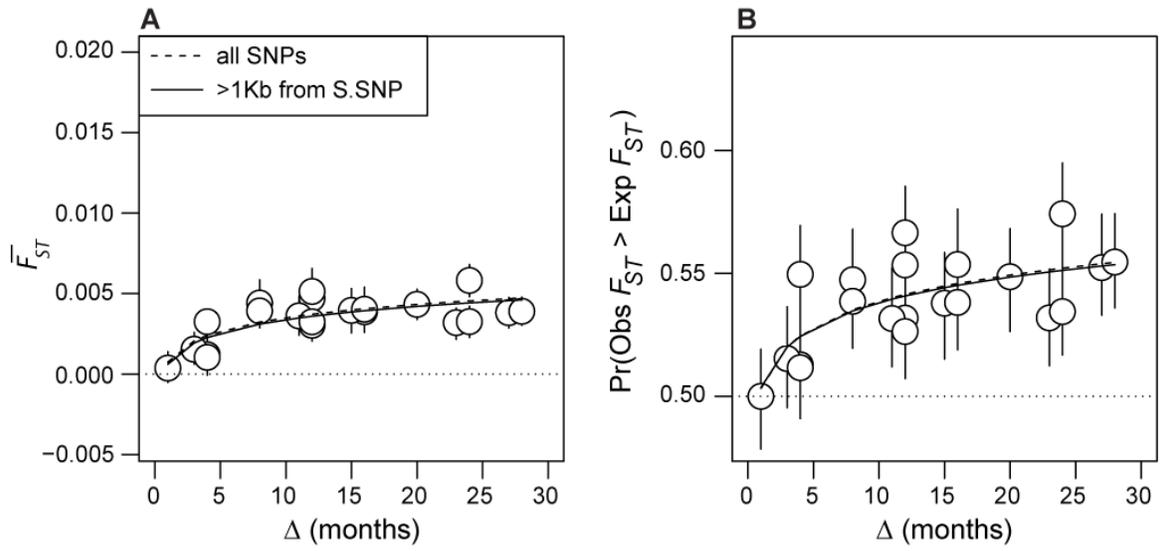

**Supplemental Figure 4.**

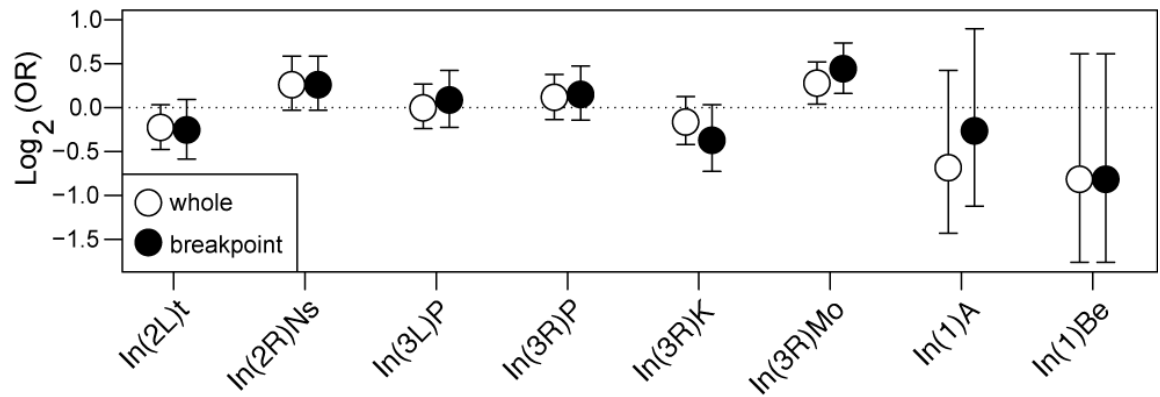

**Supplemental Figure 5.**

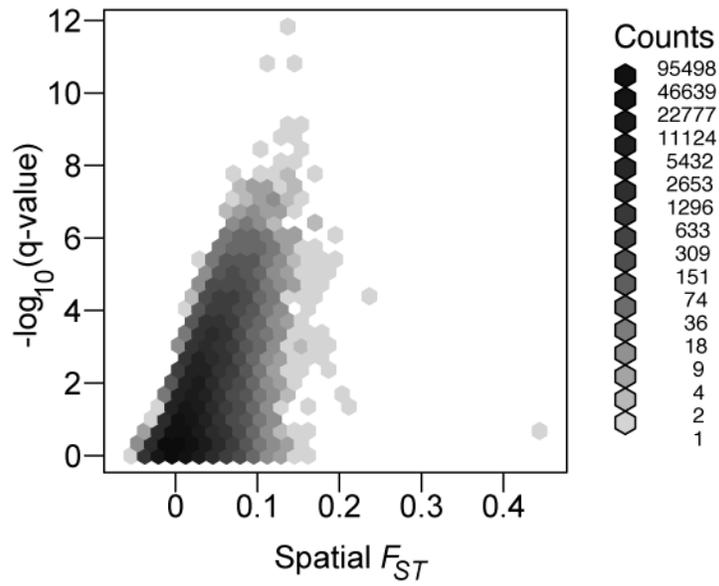

**Supplemental Figure 6.**

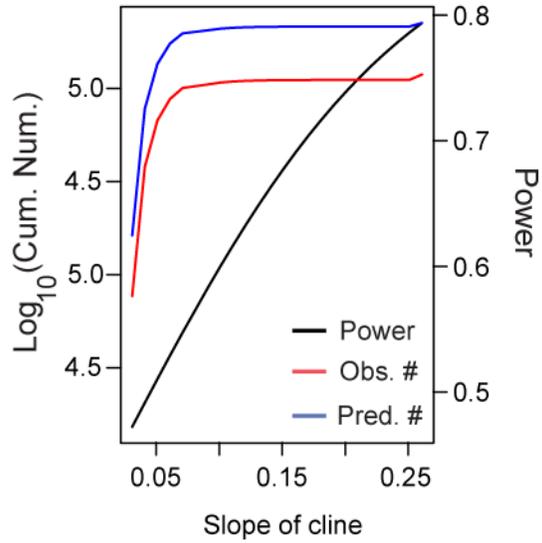

**Supplemental Figure 7.**

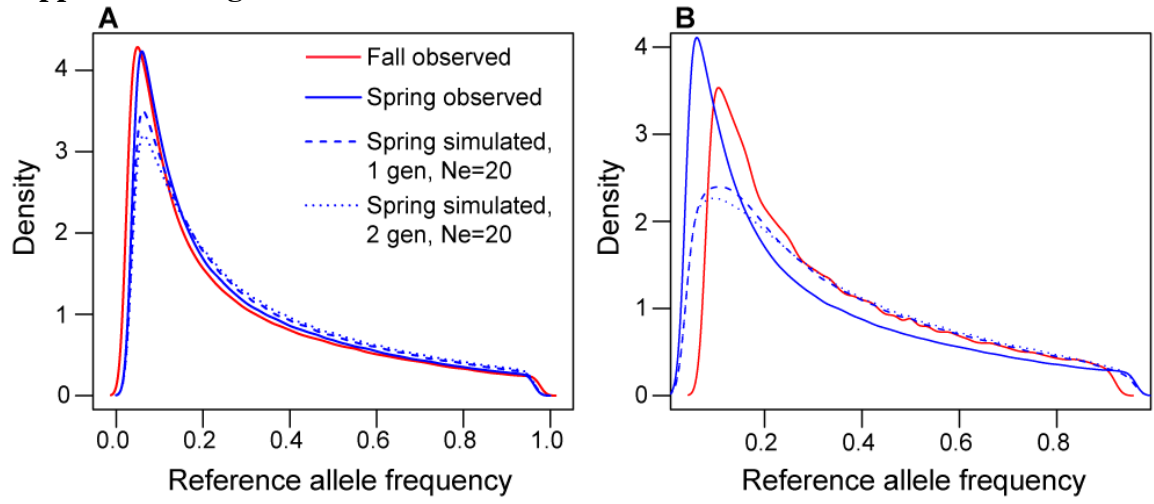

**Supplemental table 1. List of populations**

| Population | Lat. | Collection date | Chr. num (A,X)[1] | Ave. read depth[3] | SRA accession |
|---|---|---|---|---|---|
| Florida (rep 1) | 25.5 | 7/2008 & 7/2010 | 78, 39 | 69 | |
| Florida (rep 2) | 25.5 | 12/2010 | 96, 48 | 42 | |
| Georgia | 30.9 | 7/2008 | 102, 51 | 118 | |
| South Carolina | 33 | 7/2008 & 7/2010 | 96, 48 | 99 | |
| North Carolina | 35.5 | N.A. | 92, 92 | 43 | |
| Pennsylvania | 40 | 7/2009 | 110, 55 | 216 | |
| Pennsylvania | 40 | 11/2009 | 148,74 | 78 | |
| Pennsylvania | 40 | 7/2010 | 232, 116 | 29 | |
| Pennsylvania | 40 | 11/2010 | 66, 33 | 89 | |
| Pennsylvania | 40 | 7/2011 | 150, 75 | 80 | |
| Pennsylvania | 40 | 10/2011 | 94, 47 | 85 | |
| Pennsylvania | 40 | 11/2011 (post-frost) | 100,50 | 81 | |
| Maine (rep 1) | 45.5 | 10/2009 | 172, 86 | 105 | |
| Maine (rep2) | 45.5 | 10/2009 | 150, 75 | 25 | |

[1] Numbers refer to the number of autosomes (A) and sex chromosomes (X) sampled from each population
[2] See Mackay *et al.* for the provenance of the DGRP
[3] Median read depth of autosomes.

**Supplemental table 2. SNP statistics**

|  | Number of SNPs remaining after filter |
|---:|---:|
| Total identified | 2,727,167 |
| Exclude repetitive regions | 2,566,826 |
| MAF > 0.15 | 1,119,398 |
| Greater than 5 bp from indel | 1,104,461 |
| Polymorphic in DGRP | 914,959 |
| Read depth > 10X & <400X | 557,987 |
| **Total used in analysis** | **557,987** |

**Supplemental table 3. Control polymorphism factors**.

| Analysis: | Genic element | Temporal Fst decay | Inversion | Spatial Fst enrichment | Clinal q enrichment | Post-frost | Phenotype enrichment | Trans-specific | Average pairwise distance | Presence in Africa |
|---|---|---|---|---|---|---|---|---|---|---|
| Figure | 2C | 2D | 2E, S3 | 3B | 3C | 4B | 5A-D | 6A | 6B | NA |
| Qual. filter[1] | ■ | ■ | ■ | ■ | ■ | ■ | ■ | ■ | ■ | ■ |
| Chromosome | ■ | ■ | ■ | ■ | ■ | ■ | ■ | ■ | ■ | ■ |
| Rec. Rate[2] | ■ | ■ | ■ | ■ | ■ | | ■ | ■ | ■ | ■ |
| PA freq. | ■ | ■ | ■ | ■ | ■ | ■ | ■ | | ■ | ■ |
| DGRP freq. | | | | | | ■ | ■ | | | |
| African freq. | | | | | | | | ■ | | |
| Inversion North Am. freq. | | | ■ | ■ | ■ | | | | | |
| Pre – ave. delta | | | | | | ■ | | | | |
| Genic Synonymous Informative *D. sim.* reads | | | | | | | | ■ | | |
| Transspecific Common in Africa | | | | | | | | ■ | | |

[1] Qual. filter includes read depth filters, distance to indels, and presence in the DGRP as described in the Materials and Methods.

[2] Recombination rate was rounded to the nearest integer.